\documentclass[twocolumn]{article}
\usepackage[mathlines]{lineno}
\usepackage{hyperref}
\usepackage{graphicx}
\usepackage{amsfonts}
\usepackage{amsmath}
\usepackage{amssymb}
\usepackage{esint}
\usepackage{xcolor}
\usepackage{setspace}
\allowdisplaybreaks
\modulolinenumbers[5]
\usepackage{scalerel}   
\usepackage{etoolbox} 
\usepackage[resetlabels]{multibib}
\newcites{app}{Appendix literature}
\makeatletter
\modulolinenumbers[5]
\graphicspath{{figures/}}
\DeclareGraphicsExtensions{.pdf,.png}

\newcommand{\tdt}[1]{\frac{\mathrm{d} #1}{\mathrm{d} t}}
\newcommand{\dt}[1]{\frac{\partial #1}{\partial t}}
\newcommand{\dxj}[1]{\frac{\partial #1}{\partial x_j}}
\newcommand{\dxi}[1]{\frac{\partial #1}{\partial x_i}}
\newcommand{\dxk}[1]{\frac{\partial #1}{\partial x_k}}

\newcommand{\dtau}[1]{\frac{\partial #1}{\partial \tau}}

\newcommand{\dZ}[1]{\frac{\partial #1}{\partial Z}}
\newcommand{\dPsi}[1]{\frac{\partial #1}{\partial \psi}}

\newcommand{\ddZ}[1]{\frac{\partial^2 #1}{\partial Z^2}}
\newcommand{\ddPsi}[1]{\frac{\partial^2 #1}{\partial \psi^2}}

\newcommand{\dxiL}[1]{\frac{\partial}{\partial x_i}#1}

\newcommand\fav[1]{\widetilde{#1}}
\newcommand\fg{{f}'_Z}
\newcommand\ff{\widetilde{f}_Z}
\newcommand\ol[1]{\overline{#1}}
\newcommand\orho{\overline{\rho}}

\newcommand\cond[1]{\overline{\left.#1\right|\psi}}  

\newcommand\Zmin{{{Z}_{\mathrm{\scalebox{.6}{min}}} }}
\newcommand\Zmax{{{Z}_{\mathrm{\scalebox{.6}{max}}}}}
\newcommand\Zvar{\fav{{Z''}^2}}  
\newcommand\NOx{\mathrm{NO_x}}
\newcommand\iif{\fav{\textbf{F}}_Z}
\newcommand\iifb{\fav{\textbf{F}}_{Z,\beta}}

\newcommand\iiif{\fav{\textbf{I}}_Z}
\newcommand\iiifb{\fav{\textbf{I}}_{Z,\beta}}

\newcommand\avg[1]{\bigl<{#1}\bigr>}  
\newcommand\Eq[1]{Eq.~(\ref{#1})}
\newcommand\Eqs[2]{Eqs.~(\ref{#1})-(\ref{#2})}

\usepackage{authblk}
\usepackage{sectsty}
\sectionfont{\fontsize{12}{15}\selectfont}

\newcommand*\linenomathpatch[1]{%
  \cspreto{#1}{\linenomath}%
  \cspreto{#1*}{\linenomath}%
  \csappto{end#1}{\endlinenomath}%
  \csappto{end#1*}{\endlinenomath}%
}
\newcommand*\linenomathpatchAMS[1]{%
  \cspreto{#1}{\linenomathAMS}%
  \cspreto{#1*}{\linenomathAMS}%
  \csappto{end#1}{\endlinenomath}%
  \csappto{end#1*}{\endlinenomath}%
}

\expandafter\ifx\linenomath\linenomathWithnumbers
  \let\linenomathAMS\linenomathWithnumbers
  \patchcmd\linenomathAMS{\advance\postdisplaypenalty\linenopenalty}{}{}{}
\else
  \let\linenomathAMS\linenomathNonumbers
\fi

\linenomathpatch{equation}
\linenomathpatchAMS{gather}
\linenomathpatchAMS{multline}
\linenomathpatchAMS{align}
\linenomathpatchAMS{alignat}
\linenomathpatchAMS{flalign}

\begin{document}

\onecolumn
\title{Scalar mass conservation in turbulent mixture fraction based combustion models through consistent local flow parameters}

\author[1\thanks{email: \href{mailto:m.davidovic@itv.rwth-aachen.de}{m.davidovic@itv.rwth-aachen.de}}]{Marco Davidovic}
\author[1]{Heinz Pitsch}
\affil[1]{\footnotesize Institute for Combustion Technology, RWTH Aachen University, Aachen, Germany}

\maketitle


\begin{abstract}
Mixture fraction-based models are widely employed for predicting turbulent non-premixed combustion processes due to their cost-effectiveness and well-established subfilter closure. In these models, the transport of reactive scalars in physical space is decomposed into two components: scalar transport relative to mixture fraction and transport of mixture fraction in physical space. Conventional flamelet models do not consider that these two processes have to be formulated consistently, which can lead to scalar mass conservation errors. In the context of multiphase flows, scalar transport in mixture fraction space is governed by three conditional flow-dependent parameters: the conditional scalar dissipation rate, the conditional scalar diffusion rate, and the conditional spray source term. The evolution of mixture fraction in physical space is typically modeled using the presumed Filtered Density Function (FDF) approach.
This paper introduces a novel formulation for the conditional flow parameters that aligns with the presumed FDF approach, thereby ensuring scalar mass conservation. The proposed model is applied to a Large-Eddy Simulation (LES) of the inert ECN Spray A case, with a comparison against a conventional flow parameter model that employs an inverse error function shape for the scalar dissipation rate. The results indicate that the conventional model produces similar conditional dissipation rates to the new model in regions where combustion takes place. However, significant discrepancies are observed in the conditional diffusion rate, highlighting the susceptibility of the conventional model to scalar mass conservation errors for non-unity Lewis number scalars.
Lastly, the study identifies that utilizing a modified beta distribution with constant mixture fraction bounds as the FDF shape yields unrealizable conditional spray evaporation terms. Therefore, it is essential to revise the FDF model for cases in which the interaction of evaporation and chemistry plays a significant role.
\end{abstract}

%
\twocolumn
\clearpage
\section{Introduction}
Combustion of liquid fuels is widely used in technical devices such as internal combustion engines or gas turbines and, hence, provides a significant fraction of today's energy demand. However, it simultaneously contributes substantially to global warming and environmental pollution. While replacing fossil fuels with alternative liquid fuels synthesized from renewable energy and carbon feedstocks can significantly reduce net greenhouse gas emissions, harmful pollutants such as nitrogen oxides ($\NOx$) and particulate matter (PM) are still produced, and their emission need to be minimized. Although modeling of pollutant formation has been a focus of research for several decades, accurate predictions are still challenging due to the problem's complexity, which involves multi-physics that occur on a wide range of length and time scales. Large-Eddy Simulations (LES) that employ detailed-chemistry combustion models are the current state of research. Mixture fraction based models, i.e., flamelet and conditional moment closure (CMC), are most commonly used due to the predominance of non-premixed combustion in spray flames and their comparably low computational cost. Those models require conditional flow parameters, which govern the mixing of the unresolved scales, and a subfilter mixture distribution model as inputs. Even though the flow parameter and subfilter distribution models are selected independently in many studies, it can be shown that conditional flow parameters are linked to the subfilter mixture evolution, i.e., both models have to be formulated consistently to ensure conservative scalar mixing \cite{Tsai1995,Davidovic2022}. The importance of consistent transport modeling has been discussed recently in an a-priori study using DNS data \cite{Davidovic2022}. It has been shown that particularly the prediction of soot formation is prone to conservation errors if conventional flow parameter models are employed. In principle, two approaches exist for obtaining a consistent model formulation:
\begin{enumerate}
\item The conditional flow parameters can be modeled, e.g., by a presumed shape, and the mixture fraction distribution can be solved accordingly, as proposed by Ilgun, Passalacqua, and Fox \cite{Fox2020,Ilgun2021proci}. 
\item The mixture fraction distribution is modeled, and the conditional flow parameters are computed consistently.
\end{enumerate}
The first approach appears attractive since all required equations are known, and the approach always yields realizable solutions. However, it is computationally demanding to solve for the filtered density function (FDF). Additionally, the distribution function can be modeled with high certainty since a beta distribution has been found to represent turbulent scalar mixing well \cite{Cook1994,Jimenez1997,Wall2000}. 
Hence, this work focuses on the second approach. 
\par
Although analytical solutions have been derived for the conditional scalar dissipation rate in statistically homogenous turbulence \cite{OBrien1991,Tsai1998}, consistent formulations of all conditional flow parameters in inhomogeneous turbulence are not yet available. It should be noted that the wording in this paper refers to flamelet models and LES for clarity. However, the derived conditional flow parameters can also be applied to CMC models, and the model derivation can be performed similarly for the Reynolds-Averaged Navier-Stokes (RANS) approach. 
\par
The paper is structured as follows: In section \ref{sec:theory}, the consistent local flow parameter formulations are derived from the governing equations and the required modeling assumptions are presented. In the latter part, the derived model is applied to the inert ECN "Spray A" case in order to analyze and discuss the effect of the flow parameter revision on reactive model predictions. The used spray LES model and corresponding modeling assumptions are described in section \ref{sec:spray_les_model}. In section \ref{sec:application}, the spray LES model is first validated against experimental data. Afterwards, the realizability of the newly derived flow parameter model is discussed, and the flow parameter predictions are presented and compared to a conventional model. The paper closes with a summary and a conclusion in section \ref{sec:conclusion}.
\ifdefined\isthesis
\color{blue}
\else
\section{Theory and Model Derivation}\label{sec:theory}
The consistent flow parameters are derived in this section. First, the governing equations and the turbulent non-premixed spray flamelet equations are described and the required conditional flow parameters are introduced. Afterwards, a model formulation that expresses the conditional flow parameters consistently to the mixture fraction evolution is presented. The focus is directed on the model assumptions and resulting equations rather than the mathematical transformation. However, a detailed mathematical derivation is provided within the supplementary material.
\subsection{Governing Equations}
The flamelet equation for a reactive scalar, such as species mass fractions or temperature, can be derived from the balance equations for density $\rho$, mixture fraction $Z$, and the scalar $\phi$. Assuming constant Lewis numbers and Fick's law of diffusion, the governing equations can be written as
\begin{align}
\dt{\rho } + \dxi{\rho u_i } &= \dot{S}_{\rho}, \label{eq:Continuity} \\
\dt{\rho Z} + \dxi{\rho u_i Z} &= \dxi{} \left(\rho D \dxi{Z}\right) + \dot{S}_{\rho} , \label{eq:ZinSpace} \\
\dt{\rho \phi} + \dxi{\rho u_i \phi} &= \frac{1}{\mathrm{Le}_\phi}\dxi{}\left(\rho D \dxi{\phi}\right) + \dot{S}_{\rho}\phi_l + \dot{\omega}_{\phi}, \label{eq:PhiinSpace}
\end{align}
where  $t$ is the time, $x_i$ and $u_i$ are the spatial coordinate and velocity components in direction $i$, respectively, $\rho$ is the gas phase density, $D$ is the thermal diffusivity, and $\dot{\omega}_{\phi}$ is the chemical source term. The Lewis number of the mixture fraction is set to unity, while $\mathrm{Le}_\phi$ is the Lewis number of the considered scalar. When employing a Lagrangian spray model, the mass source term, $\dot{S}_{\rho}$, appears due to evaporation. $\phi_l$ denotes the scalar mass fraction in the liquid phase. 
\fi
\subsection{Turbulent spray flamelet equation}\label{ssec:turb_flamelet_equation}
In global unsteady non-premixed flamelet models, the reactive scalar is assumed to depend solely on mixture fraction and time. 
Hence, the scalar evolution can be expressed in a one-dimensional mixture fraction space, where the flamelet transformation rules are given as
\begin{align}
\dt{\phi} &\approx \dtau{\hat{\phi}}+ \dt{Z}\dZ{\hat{\phi}}\label{eq:transformation1},\\*
\dxi{\phi} &\approx \dxi{Z}\dZ{\hat{\phi}}. \label{eq:transformation2}
\end{align}
The hat symbol marks quantities defined in mixture fraction space, which are independent of the location in physical space, i.e., $\dxi{\hat{\phi}}=0$. Applying \Eqs{eq:transformation1}{eq:transformation2} to \Eqs{eq:Continuity}{eq:PhiinSpace} yields the laminar flamelet equation, which reads
\begin{equation}
\begin{split}
& \dtau{\hat{\phi}} + 
 \left[\xi_Z \left(1 - \frac{1}{\mathrm{Le}_\phi} \right) 
       + \zeta_Z\left(1-Z\right)\right] \dZ{\hat{\phi}} =   
       \ifx\isthesis\undefined \ensuremath{\\ &\qquad\qquad} \fi
   \frac{\chi_Z}{2 \mathrm{Le}_\phi} \ddZ{\hat{\phi}} 
+ \zeta_Z \left( \phi_l - \hat{\phi}\right)
+ \frac{\hat{\dot{\omega}}_{\phi}}{\hat{\rho}}. \label{eq:lam_spray_flamelet}
\end{split}
\end{equation}
The transport of $\hat{\phi}$ in mixture fraction space is governed by the local flow parameters: 
\begin{enumerate}
\item The mixture fraction dissipation rate \newline ${{\chi_Z} = 2{D \dxi{Z}\dxi{Z}}}$
\item The mixture fraction diffusion rate \newline ${{\xi_Z} = {\frac{1}{\rho}\dxi{}\left(\rho D \dxi{Z} \right)}}$
\item The normalized mass source term \newline ${{\zeta_Z} ={\frac{\dot{S}_{\rho}}{\rho}}}$
\end{enumerate}
While those parameters can be easily evaluated if the resolved flow field is known, modeling must be introduced if the flow field is unresolved, e.g., when employing the LES approach. In a global flamelet model, $\hat{\phi}$ is assumed to be independent of physical space, and thus, the local filtered scalar quantity can be obtained by
\begin{equation}
\fav{\phi}(x_i,t) = \int_{-\infty}^{\infty} \ff(\psi;x_i,t) \hat{\phi}(\psi,t) \mathrm{d}\psi,\label{eq:FDFconv}
\end{equation}
where $\ff$ denotes the local mixture fraction subfilter distribution, which is also referred to as Filtered Density Function (FDF), and $\psi$ is the mixture fraction sample space variable.
By filtering \Eqs{eq:Continuity}{eq:PhiinSpace} and invoking \Eqs{eq:transformation1}{eq:transformation2}, and \Eq{eq:FDFconv}, the global turbulent spray flamelet equation can be derived, which describes the evolution of $\hat{\phi}$ consistently to the mixture fraction subfilter evolution. A detailed derivation of the turbulent flamelet equation, including all required modeling assumptions, is provided in the supporting material. The resulting equation reads
\begin{equation}
\begin{split}
& \dtau{\hat{\phi}} + 
 \left[\xi_G \left(1 - \frac{1}{\mathrm{Le}_\phi} \right) 
       + \zeta_G\left(1-\psi\right)\right] \dPsi{\hat{\phi}} =   
       \ifx\isthesis\undefined \ensuremath{\\ &\qquad\qquad} \fi
   \frac{\chi_G}{2 \mathrm{Le}_\phi} \ddPsi{\hat{\phi}} 
+ \zeta_G \left( \phi_l - \hat{\phi}\right)
+ \frac{\hat{\dot{\omega}}_{\phi}}{\hat{\rho}}, \label{eq:turb_spray_flamelet}
\end{split}
\end{equation}
where the global conditional flow parameters are defined as
\begin{align}
\chi_G &=\frac{\int \orho \ff \cond{\chi_Z}\mathrm{d}V}{\int \orho \ff \mathrm{d}V}, 	\label{eq:globalChi}\\*
\xi_G &= \frac{\int \orho \ff \cond{\xi_Z}\mathrm{d}V}{\int \orho \ff \mathrm{d}V} ,		\label{eq:globalXi}\\*
\zeta_G &= \frac{\int \orho \ff \cond{\zeta_Z}\mathrm{d}V}{\int \orho \ff \mathrm{d}V}. \label{eq:globalZeta}
\end{align}
Note that $\orho$ is the filtered density and $V$ is the system volume. 
\subsection{Local Consistent Flow Parameter Model}\label{ssec:lcfp}
Solving the global flamelet equation (Eq.~(\ref{eq:turb_spray_flamelet})) requires knowledge of the global conditional flow parameters.  
It should be noted that even though other models exist, the conditional flow parameters are linked to the mixture evolution, i.e., transport in mixture fraction space is related to mixture fraction transport in physical space. Modeling the flow parameters independently of the mixture evolution results in scalar mass conservation errors. The global conditional dissipation and diffusion rate can be formulated consistently from the global mixture evolution for single-phase flows in closed systems \cite{Davidovic2022}. For other systems, the local parameters need to be expressed consistently to the local FDF evolution first in order to obtain the global parameters from Eqs.~(\ref{eq:globalChi})-(\ref{eq:globalZeta}).
\subsubsection{Mixture fraction FDF evolution}
The mixture fraction FDF evolution can be derived from the governing equations using the fine-grained FDF concept. Following the procedure described by Pope \cite{Pope2000}, the mixture fraction FDF transport can be obtained for flows with evaporating liquids as
\begin{equation}
\begin{split}
\ifx\isthesis\undefined \ensuremath{&} \fi
\orho\dt{\ff}+\orho \fav{u}_i \dxi{\ff} = \ifx\isthesis\undefined \else \ensuremath{&} \fi
\ifx\isthesis\undefined \ensuremath{\\&\qquad} \fi
-\dPsi{\orho\ff \cond{\xi_Z}} 
- \dxi{\orho \left(\cond{u_i} - \fav{u}_i\right) \ff} 
\ifx\isthesis\undefined \ensuremath{\\&\qquad} \else  \ensuremath{\\&}\fi
- \dPsi{}\left[\orho\ff\cond{\zeta_Z}(1-\psi)\right]
+\ff \left(\orho \cond{\zeta_Z}-\ol{\dot{S}}_{\rho}\right). \label{eq:FDFtransportExact}
\end{split}
\end{equation}
An exact relation exists between the conditional diffusion and dissipation rate that reads
\begin{equation}
-\dPsi{\orho\ff \cond{\xi_Z}} =  \dxi{}\left(\ol{\rho D\dxi{\fg}}\right) - \frac{1}{2}\ddPsi{\orho\ff\cond{\chi_Z}}, \label{eq:condDiffcondDiss1}
\end{equation}
where $\fg$ denotes the fine-grained FDF. 
Note that Eq.~(\ref{eq:FDFtransportExact}) can be used to solve for $\ff$ given models for $\cond{\chi_Z}$ and $\cond{\zeta_Z}$. This approach enforces consistent mixing in physical and flamelet space. However, solving for $\ff$ is expensive. At the same time, a good model exists for predicting subfilter mixing of conserved scalars at a much lower cost. Hence, in this work, we prescribe the FDF evolution using the presumed FDF approach and obtain the associated conditional flow parameters from Eq.~(\ref{eq:FDFtransportExact}) and Eq.~(\ref{eq:condDiffcondDiss1}). Note that although the presumed FDF approach is widely used, it does not enforce the realizability of Eq.~(\ref{eq:FDFtransportExact}), i.e., the modeled FDF evolution might require negative $\cond{\chi_Z}$ or $\cond{\zeta_Z}$ to satisfy Eq.~(\ref{eq:FDFtransportExact}). 
The presumed FDF approach prescribes a functional form of the FDF, which is parameterized by its statistical moments. 
The statistical moments are then obtained by solving modeled transport equations in physical space. 
The beta distribution has been found to approximate the turbulent mixing of conserved scalars well \cite{Cook1994,Jimenez1997,Wall2000} and is most commonly used for gaseous flows. For liquid fuel injections, the maximum mixture fraction is typically substantially below unity due to saturation effects. The distribution function should ensure a proper definition of $\ff$ in the range $\Zmin < Z < \Zmax$. For those cases, a modified beta distribution has been suggested in previous studies \cite{Reveillon2000,Ge2006} defined as
\begin{equation}
\ff =
    \begin{cases}
        \frac{1}{B(a,b)\Delta Z}{\psi^*}^{a-1} \left(1-\psi^*\right)^{b-1} & \text{if } 0 < \psi^* < 1,\\
        0 & \text{elsewhere },
    \end{cases} \label{eq:scaledBeta}
\end{equation}
with $\psi^*=\left(\psi-\Zmin\right)/\Delta Z$ and $\Delta Z=\Zmax-\Zmin$. $B(a,b)$ is the beta function for the shape parameters $a$ and $b$, which are calculated 
such that the distribution resembles the first and second mixture fraction moment
\begin{align}
a &=  \frac{\fav{Z}-\Zmin}{\Delta Z}\left[\frac{\left(\fav{Z}-\Zmin\right) \left(\Zmax - \fav{Z}\right)}{\fav{Z^2}-\fav{Z}^2} - 1\right],  \\*
b &= \frac{\Zmax-\fav{Z}}{\Delta Z}\left[\frac{\left(\fav{Z}-\Zmin\right) \left(\Zmax - \fav{Z}\right)}{\fav{Z^2}-\fav{Z}^2} - 1\right].
\end{align}
A-priori analyses have shown that the rescaled beta function is a good approximation for turbulent evaporative flows when spatially varying mixture fraction limits are employed \cite{Ge2006,Luo2011}. 
However, models for the spatial evolution of $\Zmin$ and $\Zmax$ do not yet exist, and thus, space-invariant values are typically employed.  In this study, $\Zmin=0$ is used. For $\Zmax$, it is assumed that the FDF is strongly negatively skewed for large $\fav{Z}$, and $\Zmax$ can be approximated as 
$\Zmax(t) \approx \max_{\substack{\textbf{x}\in\Omega \\ t^*\leq t}} \left[\fav{Z}(\textbf{x},t^*)\right]$, where $\Omega$ refers to the entire spatial domain.
Thus, the FDF evolution is governed solely by the first two mixture fraction subfilter moments and the spatial and temporal derivatives are given by
\begin{align}
\dt{\ff} &= \dt{\fav{Z}}\frac{\partial\ff}{\partial \fav{Z}} + \dt{\fav{Z^2}}\frac{\partial\ff}{\partial \fav{Z^2}}, \label{eq:dFDFdtmodel} \\
\dxi{\ff} &= \dxi{\fav{Z}}\frac{\partial\ff}{\partial \fav{Z}} + \dxi{\fav{Z^2}}\frac{\partial\ff}{\partial \fav{Z^2}}. \label{eq:dFDFdximodel}
\end{align}
The mixture fraction moment transport equations can be derived from Eq.~(\ref{eq:Continuity}) and Eq.~(\ref{eq:ZinSpace}) and are given in non-conservative form as
\begin{align}
\begin{split}
\ifx\isthesis\undefined \ensuremath{&} \fi
\orho\dt{ \fav{Z}} + \orho \fav{u}_i \dxi{ \fav{Z}} =
\ifx\isthesis\undefined \else\ensuremath{&} \fi
\dxi{}\left(\ol{\rho D\dxi{Z}}\right) 
\ifx\isthesis\undefined \ensuremath{\\&\quad} \fi
+ \ol{\dot{S}}_{\rho} \left(1-\fav{Z}\right) 
\ifx\isthesis\undefined\else\ensuremath{\\&}\fi
 - \dxi{}\left(\orho\fav{u_i Z} - \orho\fav{u}_i\fav{Z}\right),
\end{split}
\label{eq:Z1trans}\\
\begin{split}\label{eq:Z2trans}
\ifx\isthesis\undefined \ensuremath{&} \fi
\orho\dt{ \fav{Z^2}} + \orho \fav{u}_i \dxi{ \fav{Z^2}} =
\ifx\isthesis\undefined \else\ensuremath{&} \fi
\dxi{}\left(\ol{\rho D\dxi{Z^2}}\right) - \orho\fav{\chi}_Z 
\ifx\isthesis\undefined \ensuremath{\\&\quad} \fi
+ \ol{\dot{S}_{\rho} \left(2Z-Z^2\right)} - \ol{\dot{S}}_{\rho}\fav{Z^2} 
\ifx\isthesis\undefined\else \ensuremath{\\&} \fi
-\dxi{}\left(\orho\fav{u_i Z^2} - \orho\fav{u}_i\fav{Z^2}\right).
\end{split}
\end{align}
\subsubsection{Consistency condition}
In the flamelet formalism, the evolution of scalars in physical space is separated into two parts, the evolution of the scalar relative to the mixture fraction and the evolution of mixture fraction in physical space. Suppose at a point in space, a scalar has a given value that does not change in time and that the mixture fraction FDF is given by a delta function at a given value, but the FDF changes in time to a delta function at a different value. Then, the original value of the scalar at that point is given by
\begin{equation}
\int_{-\infty}^{\infty} \hat{\phi}(\psi,t_0) \delta\left(\psi-Z_0(x_i,t_0)\right) \mathrm{d}\psi = \fav{\phi}(x_i,t_0)
\end{equation}
 and at a later time by
\begin{equation}
\int_{-\infty}^{\infty} \hat{\phi}(\psi,t_1) \delta\left(\psi-Z_1(x_i,t_1)\right) \mathrm{d}\psi = \fav{\phi}(x_i,t_1)
\end{equation}
which are only the same if the temporal evolution of $\hat{\phi}$ is consistent with the change in $\ff$. If this is not the case, the error can, depending on the functional form of $\hat{\phi}$, be of leading order. The temporal change of $\hat{\phi}$ is given by the flamelet equations and determined by the flow parameters appearing in these equations. Hence, these have to be modeled according to the temporal evolution of the FDF.
The basis for the derivations of the conditional flow parameters is the FDF transport equation (Eq.~(\ref{eq:FDFtransportExact})) employing the partial FDF derivatives from the presumed FDF approach (Eqs.~(\ref{eq:dFDFdtmodel}) and (\ref{eq:dFDFdximodel})), where
the material derivatives of the mixture fraction moments are provided by Eqs.~(\ref{eq:Z1trans}) and (\ref{eq:Z2trans}). Note that the subfilter moment fluxes need to be formulated consistently to the subfilter FDF fluxes so that the moments described by both moment equations and pdf transport equation are the same.

When the subfilter FDF flux is modeled using a gradient transport approach, i.e., $\left(\cond{u_i} - \fav{u}_i\right) \ff = - D_{t,\psi} \dxi{\ff}$,
where $D_{t,\psi}$ denotes the eddy-diffusivity at sample space location $\psi$, consistent subfilter moment fluxes can be easily found if the eddy-diffusivity is assumed to be independent of $\psi$, i.e., $D_{t,\psi}=\fav{D}_t$. The consistent subfilter moment fluxes are then given by $\fav{u_i Z} - \fav{u}_i\fav{Z}  = - \fav{D}_{t} \dxi{\fav{Z}}$ and $\fav{u_i Z^2} - \fav{u}_i\fav{Z^2}  = - \fav{D}_{t} \dxi{\fav{Z^2}}$.
\par
The filtered molecular diffusion fluxes can be expressed consistently by assuming $\dxi{\orho \fav{D}} = \orho\cond{\frac{1}{\rho}\dxi{\rho D}}$, which yields the filtered molecular fluxes of the $\ff$, $\fav{Z}$, and $\fav{Z^2}$ as $\ol{\rho D\dxi{\ff}}  = \orho\fav{D}_{t} \dxi{\ff}$, $\ol{\rho D\dxi{Z}}  = \orho\fav{D}_{t} \dxi{\fav{Z}}$, and $\ol{\rho D\dxi{Z^2}}  = \orho\fav{D}_{t} \dxi{\fav{Z^2}}$, respectively.
Note that the corresponding mathematical proofs are provided in the supplementary material. 
\par
With these assumptions, a consistency condition can be derived between the flow parameters appearing in the flamelet equations and the FDF that reads
\begin{equation}
\begin{split}
&
\frac{1}{2}\ddPsi{\orho\ff\cond{\chi_Z}} + \dPsi{}\left[\orho\ff\cond{\zeta_Z}(1-\psi)\right] 
-\ff\orho \cond{\zeta_Z} \\&=  
\frac{\orho \fav{\chi}_{\fav{Z}\fav{Z}}}{2} \frac{\partial^2\ff}{\partial \fav{Z}^2}  
+\orho \fav{\chi}_{\fav{Z}\fav{Z^2}}  \frac{\partial^2\ff}{\partial\fav{Z}\partial \fav{Z^2}}
+ \frac{\orho \fav{\chi}_{\fav{Z^2}\fav{Z^2}}}{2} \frac{\partial^2\ff}{\partial \fav{Z^2}^2}
\\&\quad
-\ff\ol{\dot{S}}_{\rho}-\left[\ol{\dot{S}}_{\rho} \left(1-\fav{Z}\right)\right]\frac{\partial\ff}{\partial{\fav{Z}}} 
\\&\quad
+\left(\orho\fav{\chi}_Z - \ol{\dot{S}_{\rho} \left(2Z-Z^2\right)}  + \ol{\dot{S}}_{\rho}\fav{Z^2} \right)\frac{\partial\ff}{\partial{\fav{Z^2}}} 
, \label{eq:condFlowBasis2}
\end{split}
\end{equation}
with 
\begin{align}
\fav{\chi}_{\fav{Z}\fav{Z}} &=  2\left(\fav{D} + \fav{D}_t\right) \dxi{\fav{Z}}\dxi{\fav{Z}}, \\
\fav{\chi}_{\fav{Z}\fav{Z^2}} &= 2 \left(\fav{D} + \fav{D}_t\right) \dxi{\fav{Z}}\dxi{\fav{Z^2}}, \\
\fav{\chi}_{\fav{Z^2}\fav{Z^2}} &= 2 \left(\fav{D} + \fav{D}_t\right) \dxi{\fav{Z^2}}\dxi{\fav{Z^2}}.
\end{align}
\subsubsection{Conditional mixture fraction dissipation rate}
The conditional dissipation rate can be derived by considering that the instantaneous scalar dissipation rate of a scalar mixing field is defined and hence directly given by the local mixture fraction gradient and the mixture fraction diffusivity. It is, therefore, only indirectly affected by mass source terms. It is clear that the presence of mass source terms affects the scalar dissipation rate evolution by increasing or reducing scalar gradients over time. However, since the instantaneous conditional dissipation rate is of interest for solving the unsteady flamelet equations, all mass source term effects can be neglected in Eq.~(\ref{eq:condFlowBasis2})  for deriving $\cond{\chi_Z}$, i.e., $\dot{S}_\rho=0$ and thus $\cond{\zeta_Z}=0$. Hence, $\cond{\chi_Z}$ is obtained by integrating Eq.~(\ref{eq:condFlowBasis2}) twice in mixture fraction sample space. The integration can be easily performed since all variables, apart from $\ff$, are independent of $\psi$, and can therefore be factorized from the integral. Additionally, the order of integration and differentiation can be altered since the integration bounds are independent of the mixture fraction moments. Hence, the conditional mixture fraction dissipation can be obtained via
\begin{equation}
\begin{split}
\ff\cond{\chi_Z} = & 2\fav{\chi}_Z \frac{\partial\iif}{\partial{\fav{Z^2}}}
+
\fav{\chi}_{\fav{Z}\fav{Z}}\frac{\partial^2\iif}{\partial \fav{Z}^2}  
+2\fav{\chi}_{\fav{Z}\fav{Z^2}}\frac{\partial^2\iif}{\partial\fav{Z}\partial \fav{Z^2}} 
\\&+\fav{\chi}_{\fav{Z^2}\fav{Z^2}}\frac{\partial^2\iif}{\partial \fav{Z^2}^2} .
\label{eq:condCHIexact}
\end{split}
\end{equation}
A new function has been introduced for better readability, which is defined by 
${\iif  \left(\psi;\mathbf{x},t\right) = \iint_{-\infty}^\psi \ff\left(\psi';\fav{Z},\fav{Z^2}\right)\mathrm{d}\psi'\mathrm{d}\psi'}$.
For the modified beta distribution given in Eq.~(\ref{eq:scaledBeta}), $\iif$ is given by ${\iifb =\frac{\Delta Z}{B(a,b)}  \left[\psi^* B_{{\psi}^{*}}(a,b)  - B_{{\psi}^{*}}(a+1,b)\right]}$, where $B_{{\psi}^{*}}(a,b)$ is the incomplete beta function at ${\psi}^{*}$. $B_{{\psi}^{*}}(a,b)$ can be numerically evaluated using a continued fraction method \cite{Cuyt2008}, which converges rapidly for sufficiently large variances. At the same time, the convergence rate is slow for tiny variances. Fortunately, the beta distribution can be well approximated by a delta distribution for those cases. Hence, it is computationally more favorable to switch to a delta distribution when the beta distribution is not sufficiently resolved by the sample space grid. For a delta distribution, an analytic solution exists for Eq.~(\ref{eq:condCHIexact}) that reads
$\ff\cond{\chi_Z} =  \ff2\left(\fav{D} + \fav{D}_t\right)\dxi{\fav{Z}}\dxi{\fav{Z}}$.
In this study, we switch from a beta distribution to a delta distribution if the variance falls below $5\cdot10^{-6}$. 
\par
Note that Eq.~(\ref{eq:condCHIexact}) contains both the conditionally and unconditionally filtered scalar dissipation rate. Eq.~(\ref{eq:condCHIexact}) can be integrated over the entire mixture fraction sample space in an attempt to solve for $\fav{\chi}_Z$, yielding
\begin{equation}
\begin{split}
\fav{\chi}_Z =& 
2\fav{\chi}_Z \frac{\partial\iiif}{\partial{\fav{Z^2}}} 
 +
\fav{\chi}_{\fav{Z}\fav{Z}t}\frac{\partial^2\iiif}{\partial \fav{Z}^2}  
+2\fav{\chi}_{\fav{Z}\fav{Z^2}t}\frac{\partial^2\iiif}{\partial\fav{Z}\partial \fav{Z^2}} 
\\&+\fav{\chi}_{\fav{Z^2}\fav{Z^2}t}\frac{\partial^2\iiif}{\partial \fav{Z^2}^2} 
,\label{eq:meanCHIexact}
\end{split}
\end{equation}
where $\iiif= \int_{-\infty}^{\infty} \iif\left(\psi;\fav{Z},\fav{Z^2}\right) \mathrm{d}\psi$ has been introduced. It is interesting to note that using the modified beta function as FDF, $\iiif$ reduces to a simple analytic function $\iiifb = 1/2\left({\Zmax^2-2\fav{Z}\Zmax+\fav{Z^2}}\right)$.
Hence, all second partial derivatives of $\iiifb$ that are present in Eq.~(\ref{eq:meanCHIexact}) vanish. More importantly, $\frac{\partial\iiifb}{\partial{\fav{Z^2}}}=\frac{1}{2}$ and hence, Eq.~(\ref{eq:meanCHIexact}) is always satisfied. This means that the filtered dissipation rate is an external parameter if a modified beta distribution is employed and cannot be obtained from Eq.~(\ref{eq:condCHIexact}).
However, if a delta distribution is employed, the mean scalar dissipation rate is determined by Eq.~(\ref{eq:meanCHIexact}) and reads $\fav{\chi}_{Z,\delta}=2\left(\fav{D}+\fav{D}_t\right)\dxi{\fav{Z}}\dxi{\fav{Z}}$.
It should be noted that $\fav{D}_t=0$ and $\fav{Z}=Z$ in the case of zero subfilter variance; hence, the scalar dissipation rate definition is recovered.
\subsubsection{Conditional mixture fraction diffusion rate}
The conditional diffusion rate can be determined from the conditional dissipation rate and Eq.~(\ref{eq:condDiffcondDiss1}) and reads
\begin{equation}
\begin{split}
\ff\cond{\xi_Z} =& 
       -\xi_{\fav{Z}}						\frac{\partial\fav{F}_Z}{\partial \fav{Z}}  
-\left(\xi_{\fav{Z^2}} - \fav{\chi}_Z\right)		\frac{\partial\fav{F}_Z}{\partial \fav{Z^2}} 
\\&
+ \frac{\fav{\chi}_{\fav{Z}\fav{Z},t}}{2}		\frac{\partial^2\fav{F}_Z }{\partial \fav{Z}^2}  
+         \fav{\chi}_{\fav{Z}\fav{Z^2},t}		\frac{\partial^2\fav{F}_Z}{\partial \fav{Z}\fav{Z^2}}  
\\&
+ \frac{\fav{\chi}_{\fav{Z^2}\fav{Z^2},t}}{2}  \frac{\partial^2\fav{F}_Z}{\partial \fav{Z^2}^2}
,\label{eq:xiMdlwsrc}
\end{split} 
\end{equation}
with
\begin{align}
\xi_{\fav{Z}} 				&= \frac{1}{\orho}\dxi{}\left(\orho \fav{D}\dxi{\fav{Z}}\right), \\
\xi_{\fav{Z^2}} 				&= \frac{1}{\orho}\dxi{}\left(\orho \fav{D}\dxi{\fav{Z^2}}\right), \\
\fav{\chi}_{\fav{Z}\fav{Z},t} 		&=  2\fav{D}_t   \dxi{\fav{Z}}\dxi{\fav{Z}}, \\
\fav{\chi}_{\fav{Z}\fav{Z^2},t} 	&= 2 \fav{D}_t   \dxi{\fav{Z}}\dxi{\fav{Z^2}}, \\
\fav{\chi}_{\fav{Z^2}\fav{Z^2},t}	&= 2 \fav{D}_t   \dxi{\fav{Z^2}}\dxi{\fav{Z^2}}.
\end{align}
Note that $\fav{F}_Z$ refers to the cumulative filtered density function (CDF), which is defined as $\fav{F}_Z=\int_{-\infty}^\psi \ff\left(\psi';\fav{Z},\fav{Z^2}\right)\mathrm{d}\psi'$. For a modified beta distribution, $\fav{F}_Z$ is given by 
$\fav{F}_{Z,\beta} =\frac{B_{{\psi}^{*}}(a,b)}{B(a,b)}$.
Similarly to $\ff\cond{\chi_Z}$, an analytical solution can be found for $\ff\cond{\xi_Z}$ if a delta distribution is employed, which reads
$\ff\cond{\xi_Z} = \ff\frac{1}{\orho}\dxi{} \left(\orho \fav{D} \dxi{\fav{Z}}\right)$.
\subsubsection{Conditional normalized source term}
The conditional normalized mass source term can be obtained by inserting Eq.~(\ref{eq:condCHIexact}) into Eq.~(\ref{eq:condFlowBasis2}), which results in 
\begin{equation}
\begin{split}
\ifx\isthesis\undefined\ensuremath{&} \fi
(\psi-1)\dPsi{}\left(\orho\ff\cond{\zeta_Z}\right)  + 2 \ff \orho \cond{\zeta_Z} =  
\ifx\isthesis\undefined\else\ensuremath{&} \fi
\ol{\dot{S}}_{\rho}\ff 
\ifx\isthesis\undefined\ensuremath{\\&} \fi
+\ol{\dot{S}}_{\rho} \left(1-\fav{Z}\right) \frac{\partial\ff}{\partial \fav{Z}} 
\ifx\isthesis\undefined\else\ensuremath{\\&} \fi
+\left[ \ol{\dot{S}_{\rho} \left(2Z-Z^2\right)}  - \ol{\dot{S}}_{\rho}\fav{Z^2}\right]\frac{\partial\ff}{\partial \fav{Z^2}}  . \label{eq:srcODE}
\end{split}
\end{equation}
For a given set of mixture fraction moments, Eq.~(\ref{eq:srcODE}) resembles a first-order linear ordinary differential equation (ODE) that relates the conditional spray source term to the presumed FDF shape and mixture fraction moment source terms. An analytical solution for Eq.~(\ref{eq:srcODE}) can be found using the method of integrating factors, resulting in
\begin{equation}
\begin{split}
\ff \cond{\zeta_Z} =& \frac{\ol{\dot{S}}_{\rho}}{\orho\left(\psi-1\right)^2} \fav{F}_Z^* +
  \frac{\ol{\dot{S}}_{\rho} \left(1-\fav{Z}\right) }{\orho\left(\psi-1\right)^2} 	   \frac{\partial \fav{F}_Z^*}{\partial \fav{Z}} 
\ifx\isthesis\undefined\ensuremath{\\&} \fi
  +\frac{\ol{\dot{S}_{\rho} \left(2Z-Z^2\right)} -\ol{\dot{S}}_{\rho}\fav{Z^2} }{\orho\left(\psi-1\right)^2} \frac{\partial\fav{F}_Z^* }{\partial \fav{Z^2}}. 
\end{split}\label{eq:condSrc} 
\end{equation}
Again, a new function $\fav{F}_Z^* = \int_{-\infty}^\psi(\psi'-1) \ff\mathrm{d} \psi' $ has been introduced for better readability. If a modified beta distribution is employed, $\fav{F}_Z^*$ is given by $\fav{F}^*_{Z,\beta} = \frac{\Delta Z }{B\left(a,b\right)}\left[B_{\psi^*}\left(a+1,b\right) - \frac{1-\Zmin}{\Delta Z}B_{\psi^*}\left(a,b\right) \right]$,
while for a delta distribution Eq.~(\ref{eq:condSrc}) reduces to 
$\ff\cond{\zeta_Z} = \ff\frac{\ol{\dot{S}}_{\rho}}{\orho}$.  
\par
The derived model equations will be referred to as the local consistent flow parameter (LCFP) model in the following.
\ifdefined\isthesis
\color{black}
\fi

\ifdefined\isthesis
\color{blue}
\subsection{Filtered Governing Equations}
Although the fluid motion is given by \Eqs{eq:Continuity}{eq:Energy},
\else
\section{Spray LES Model}\label{sec:spray_les_model}
The derived flow parameter model will be applied to a high-pressure fuel injection under engine-relevant conditions in the latter part of this paper. The spray and flow models used for those simulations are presented in this section.
\subsection{Filtered Governing Equations}
The flow of the continuous fluid phase is governed by the balance of mass, momentum, and energy, expressed by the Navier-Stokes equations. Even though this set of equations is valid for turbulent flows,
\fi
obtaining a direct numerical solution for high Reynolds number flows is typically not feasible since the broad range of scales cannot be resolved due to limited computational resources. In LES, the governing equations are filtered; thus, only the large-scale turbulent structures need to be resolved, hence reducing the computational effort substantially.
\ifdefined\isthesis
\color{black}
\par
For reactive flows, density-weighted filtering, i.e., Favre filtering, is commonly employed, which can be achieved by integrating the flow variables, e.g., a scalar $\phi$, against a filter kernel, $\mathcal{F}$, according to
\begin{align}
\orho(\mathbf{x},t) \fav{\phi}(\mathbf{x},t) = \ol{\rho(\mathbf{x},t) \phi(\mathbf{x},t)} &= \int_V \mathcal{F}\left(\mathbf{r}\right)\rho(\mathbf{x}+\mathbf{r}) \phi(\mathbf{x}+\mathbf{r})  \mathrm{d}\mathbf{r}, \label{sm_eq:favre_filtering} \\
\orho(\mathbf{x},t) &= \int_V \mathcal{F}\left(\mathbf{r}\right)\rho(\mathbf{x}+\mathbf{r})   \mathrm{d}\mathbf{r}.
\end{align}
To derive the balance equations for the filtered quantities, commutation of filtering and differentiation has to be assumed, which is not universally valid \cite{Ghosal1995} but is commonly made \cite{Poinsot2005} yielding
\color{blue}
\else
Assuming unity Lewis number for all species, the filtered balance equations read
\fi
\begin{align}
\dt{\bar{\rho}} + \dxi{\bar{\rho}\fav{u}_i} =&  \ol{\dot{S}}_\rho, \label{eq:ContinuityFiltered} 
\\*
\begin{split}
\dt{\bar{\rho}\fav{u}_j} + \dxi{\bar{\rho}\fav{u}_i\fav{u}_j} =&  \dxiL{\left[\ol{\sigma}_{ij}-\ol{\rho}\left(\fav{u_i u_j}-\fav{u}_i \fav{u}_j\right)\right]} 
\ifx\isthesis\undefined \ensuremath{\\&} \fi
+ \ol{\dot{S}}_{u_j} \label{eq:MomentumFiltered} ,
\end{split}
\\*
\begin{split}
\dt{\bar{\rho}\fav{e}}+\dxi{\bar{\rho}\fav{u}_i\fav{e}} =& \dxiL{ \left[\orho\fav{D \dxi{h}}-\ol{\rho}\left(\fav{u_i e} - \fav{u_i}\fav{e} \right)\right]}  
\ifx\isthesis\undefined \ensuremath{\\&} \fi
+ \ol{\sigma_{ij}\dxj{u_i}} + \ol{\dot{S}}_e
\ifx\isthesis\undefined\else
+\ol{\dot{Q}}
\fi
,\label{eq:EnergyFiltered}
\end{split} 
\end{align}
\ifx\isthesis\undefined
where $\sigma_{ij}$ is the stress tensor, $e$ the energy, $h$ the enthalpy, and $T$ the temperature. Contributions from the liquid phase are included in the volumetric mass source, $\dot{S}_\rho$, momentum source, $\dot{S}_{u_i}$, and energy source, $\dot{S}_{e}$. 
The stress tensor of a Newtonian fluid is defined as $\sigma_{ij}=2\rho\nu\left(S_{ij}-\frac{1}{3}\delta_{ij}\dxk{u_k}\right)-P\delta_{ij}$, where $\nu$ is the kinematic viscosity and $P$ is the pressure. $\delta_{ij}$ is the Kronecker delta, and $S_{ij}$ is the strain-rate defined as $S_{ij}=\frac{1}{2}\left(\dxj{u_i}+\dxi{u_j}\right)$.
The ideal gas law is employed as equation of state in this work. The filtered equation reads
\else
\color{black}
Additionally, the filtered ideal gas law and energy-temperature relation are given as
\color{blue}
\fi
\begin{equation}
\bar{P}=\bar{\rho}\mathcal{R} \fav{\left(\frac{T}{W}\right)}, \label{eq:filidealGas}
\end{equation}
\ifx\isthesis\undefined
where $\mathcal{R}$ is the universal gas constant and $W$ the molecular weight of the gas mixture. The temperature depends on the energy and mixture composition according to
\else
\color{black}
and
\color{blue}
\fi
\begin{equation}
\fav{e} = \frac{1}{\orho}\sum_{\alpha=1}^{n_\alpha} [\ol{\rho{Y}_\alpha h_\alpha(T)}]-\frac{\ol{P}}{\orho}
\ifx\isthesis\undefined,\else.\fi 
\label{eq:filTfromE}
\end{equation}
\ifx\isthesis\undefined
where $Y_\alpha$ is mass fraction and $h_\alpha$ the specific enthalpy of species $\alpha$. 
\else
\color{black}
\color{blue}
\fi

\subsection{Subfilter Closure}\label{sec:subfilter_flow}
Eqs.~(\ref{eq:ContinuityFiltered})-(\ref{eq:EnergyFiltered}) contain unclosed subfilter stresses and fluxes, which arise from the non-linear convection terms, and unclosed filtered laminar fluxes and sources. Furthermore, Eqs.~(\ref{eq:filidealGas}) and (\ref{eq:filTfromE}) are also unclosed. The closure approximations used in this work are described in the following. 
\par
The unresolved Reynolds stresses $(\fav{u_i u_j} - \fav{u}_i \fav{u}_j)$ are modeled using a dynamic Smagorinsky model \cite{Germano1991} with averaging along Lagrangian trajectories \cite{Meneveau1996}, while the filtered molecular viscous stress is modeled by assuming that $\nu$ and all velocity gradients are uncorrelated resulting in $\ol{\sigma}_{ij} = 2\orho\fav{\nu} \left(\fav{S}_{ij}-\frac{1}{3}\delta_{ij}\dxk{\fav{u}_k}\right)$. 
\par
The turbulent fluxes in \Eq{eq:EnergyFiltered} can be simplified by neglecting subfilter pressure fluctuations to
\begin{equation}
\begin{split}
\ol{\rho}\left(\fav{u_i e} - \fav{u_i}\fav{e} \right) =& \ol{\rho}\left(\fav{u_i h} - \fav{u_i}\fav{h} \right) - \left(\ol{u_i P} - \ol{u_i}\ol{P} \right) 
\\=& \ol{\rho}\left(\fav{u_i h} - \fav{u_i}\fav{h} \right). \label{eq:eh_flux}
\end{split}
\end{equation}
The enthalpy flux includes both heat conduction and enthalpy transport due to species diffusion. Species diffusion in physical space is governed by mixture fraction transport in physical space and species transport in mixture fraction space when a flamelet model is employed. Thus, to prevent decoupling of mixture fraction and energy, it is preferable to express the enthalpy and mixture fraction flux consistently. 
\par
\ifx\isthesis\undefined\else
\textcolor{red}{TODO}\\
\fi
The total differential of enthalpy is
\begin{equation}
\mathrm{d}h = c_p\mathrm{d}T + \sum_{\alpha=0}^{n_\alpha} h_{\alpha}\mathrm{d}Y_{\alpha}. \label{eq:totDiffH}
\end{equation}
Note that the local species mass fractions are governed by the flamelet solution, i.e., $Y_{\alpha}=\hat{Y}_\alpha$, while the flamelet temperature, $\hat{T}$, represents a conditionally volume-averaged temperature. Note that $\hat{T}$ can deviate from the local temperature due to droplet heating or compressibility. Those deviations are expressed via a temperature correction variable, $\Delta T$, defined by $T=\hat{T}+\Delta T$. Then, Eq.~(\ref{eq:totDiffH}) can be expressed as 
\begin{equation}
\mathrm{d}h = \mathrm{d}\hat{h} + c_p\mathrm{d}\Delta T + \sum_{\alpha=0}^{n_\alpha} \Delta h \mathrm{d}\hat{Y}_\alpha, \label{eq:totDiffDeltaH}
\end{equation} 
where $\Delta h = \int_{\hat{T}}^{\hat{T}+\Delta T} c_{p,\alpha}\mathrm{d}T$. The last term in \Eq{eq:totDiffDeltaH} is usually small and will be neglected in the following. 
\par
Employing a gradient diffusion model for the turbulent fluxes in combination with \Eq{eq:eh_flux} and \Eq{eq:totDiffDeltaH}, the filtered diffusive and turbulent subfilter fluxes in Eq.~(\ref{eq:EnergyFiltered}) can be simplified to
\begin{equation}
\begin{split}
&\orho\fav{D \dxi{h}}-\ol{\rho}\left(\fav{u_i e} - \fav{u_i}\fav{e} \right) = 
\ifx\isthesis\undefined \ensuremath{\\&\qquad} \fi
\orho\fav{D \dxi{\hat{h}}}+\overline{\lambda \dxi{\Delta T}}-\ol{\rho}\left(\fav{u_i h} - \fav{u_i}\fav{h} \right).
\end{split}
\end{equation}
A consistent subfilter energy flux model can then be formulated as
\begin{equation}
\begin{split}
& \orho\fav{D \dxi{\hat{h}}}+\overline{\lambda \dxi{\Delta T}}-\ol{\rho}\left(\fav{u_i h} - \fav{u_i}\fav{h} \right)=
\\&\qquad
\orho\left(\fav{D}+\fav{D}_t\right) \dxi{\hat{h}} + \left(\overline{\lambda} + \overline{\lambda}_t\right) \dxi{\fav{\Delta T}}, \label{eq:sgsEflux}
\end{split}
\end{equation}
where ${\fav{\Delta T} =  \fav{T}-\fav{\hat{T}}}$ with  $\fav{\hat{T}}=\int_{-\infty}^\infty  \ff \hat{T} \mathrm{d}\psi$. 
In \Eq{eq:sgsEflux}, the right-hand side comprises two distinct terms. The first term encompasses the energy flux associated with the evolution of the mixture fraction field. The second term, on the other hand, accounts for additional heat conduction, which arises from local variations in temperature between the flamelet and the flow solution, which may result from compressibility or droplet heating. To maintain the consistency between the mixture fraction and energy fields, $\fav{D}_t$ in \Eq{eq:sgsEflux} must be equal to the eddy-diffusivity applied in the mixture fraction moment equation. The eddy-conductivity, $\ol{\lambda}_t$, is obtained by dynamically evaluating the eddy-diffusivity, $\fav{D}_{t,\Delta T}$, for $\fav{\Delta T}$. Then, $\fav{D}_{t,\Delta T}$ is converted to a conductivity, i.e., $\ol{\lambda}_t =\fav{c_p} \orho \fav{D}_{t,\Delta T}$.
The remaining filtered energy source terms are modeled as $\ol{P\dxi{u_i}}=\ol{P}\dxi{\fav{u}_i}$, and $\ol{\sigma_{ij}\dxj{u_i}}=\ol{\sigma}_{ij}\dxj{\fav{u}_i}$. 
\par
For closure of \Eq{eq:filidealGas} and \Eq{eq:filTfromE}, it should be noted that the mixture composition and the temperature are strongly correlated in spray combustion cases when cold fuel is injected into a hot atmosphere. In order to consider those correlations, \Eq{eq:filidealGas} and \Eq{eq:filTfromE} can first be written as
\begin{align}
\ol{P}  &=  \orho\mathcal{R}\frac{\fav{T}}{\fav{W}} + \Delta\ol{P}_{sgs} , \\*
\fav{e} &= \fav{h}-\frac{\ol{P}}{\orho} = \sum_{\alpha=0}^{n_\alpha} \left[\fav{Y}_\alpha h_\alpha(\fav{T})\right]+ \Delta\fav{h}_{sgs} - \frac{\ol{P}}{\orho},
\end{align}
where $\Delta\ol{P}_{sgs}$ and $\Delta\fav{h}_{sgs}$ contain the subfilter correlations of $T$ and $W$, and $h_\alpha(T)$ and $Y_\alpha$, respectively. When a flamelet model is employed, the subfilter distribution of $W$ and $Y_\alpha$ depends entirely on $Z$. Although $T$ might locally vary from the flamelet solution $\hat{T}$ due to pressure waves or droplet heating, the temperature fluctuations caused by mixture field fluctuations are usually predominant. Hence, it seems reasonable to approximate the temperature subfilter fluctuations using the flamelet solution and the local mixture fraction FDF. The correlation correction terms are then modeled as
\begin{align}
\Delta \ol{P}_{sgs} &=\orho\mathcal{R}\left(\int_{-\infty}^\infty  \ff \frac{\hat{T}}{\hat{W}}\mathrm{d}\psi - \frac{\fav{\hat{T}}}{\fav{W}}\right), \\
\Delta\fav{h}_{sgs} &= \sum_{\alpha=0}^{n_\alpha} \left[ \int_{-\infty}^\infty  \ff \hat{Y}_\alpha h_\alpha(\hat{T})\mathrm{d}\psi  - \fav{Y}_\alpha h_\alpha\left( \fav{\hat{T}}\right)\right],
\end{align}
where $\fav{\hat{T}}=\int_{-\infty}^\infty  \ff \hat{T} \mathrm{d}\psi $.
\par
\ifx\isthesis\undefined
Besides the flow field, the first two subfilter mixture fraction moments are solved according to Eqs.~(\ref{eq:Z1trans}) and (\ref{eq:Z2trans}). The turbulent fluxes are closed using a Smagorinsiky-type model, i.e., $\fav{D}_t = 2 C_Z\ol{\Delta}^2|\fav{S}_{ij}|$. $C_Z$ is evaluated dynamically for the $\fav{Z}$ flux \cite{Pierce1998} using Lagrangian averaging \cite{Meneveau1996}. $\ol{\Delta}$ denotes the filter width. 
The second moment equation includes $\fav{\chi}_Z$ as sink term, which additionally requires closure. $\fav{\chi}_Z$ can be written in terms of the resolved scales and an unresolved part, $\fav{\chi}_{Z,sgs}$, according to
\begin{equation}
\fav{\chi}_Z = 2\fav{D}\dxi{\fav{Z}}\dxi{\fav{Z}} + \fav{\chi}_{Z,sgs}.
\end{equation}
The subfilter scalar dissipation rate is modeled using a mixing time-scale relation \cite{Peters2000,Raman2006} yielding
\begin{equation}
\fav{\chi}_{Z,sgs}=C_\chi \fav{D}_t \left(\fav{Z^2}-\fav{Z}^2\right).
\end{equation}
The values of $C_\chi$ vary in the literature between approximately $2<C_\chi<8$ \cite{Peters2000,Ihme2008b,Kaul2011,De2013}. In this study, a value of $C_\chi=4$ is used.
\fi

\ifdefined\isthesis
\color{blue}
\subsection{Liquid Phase Model}
\else
\subsection{Liquid Phase Model Equations}
\fi
High-pressure injection events are characterized by high Reynolds and Weber numbers. Hence, during the primary breakup process, the liquid fuel jet disintegrates rapidly into small and mostly spherical structures. Those droplets may break up further into even smaller droplets (secondary breakup) and evaporate. Resolving the liquid structures in such an injection event is computationally unreasonable. Thus, as done in this work, Lagrangian particle models are commonly applied. In the Lagrangian Particle Tracking (LPT) approach, the shape of the droplets is presumed, and the droplet quantities are integrated along their trajectory in time. Interactions with the surrounding gas need to be modeled. In the present study, standard drag, evaporation, and breakup models are applied, which are presented briefly in the following.
\par
\ifdefined\isthesis
\subsubsection{Droplet Drag and Evaporation Model}
\fi
If the resolved flow field is known, the droplet position, $x_{d,j}$, and velocity, $u_{d,j}$ can be solved according to Miller and Bellan \cite{Miller1999} using
\begin{align}
\tdt{x_{d,j}}=&u_{d,j} ,\label{eq:xDres} \\
\tdt{u_{d,j}}=&\frac{f_1}{\tau_d}\left({u}_j - u_{d,j}\right), \label{eq:uDres}
\end{align}
where $\tau_d=\rho_d d^2_d/(18\mu_r)$ is the particle time constant with $\mu_r$ denoting the dynamic viscosity of the surrounding gas and $f_1$ is an empirical correction factor that accounts for finite droplet Reynolds numbers. The droplet mass, $m_{d}$, and temperature, $T_d$, are solved using the evaporation model by Miller and Bellan \cite{Miller1999}
\begin{align}
\tdt{m_d} &= -\frac{\mathrm{Sh}}{3\mathrm{Sc}}\frac{m_d}{\tau_d} \Pi_d \label{eq:evapMres}, \\
\tdt{T_d}& = \frac{\dot{Q}_d}{c_l} + \frac{1}{m_d}\tdt{m_d}\frac{L_v}{c_l} \label{eq:evapTres},
\end{align}
where $\dot{Q}_d = \frac{\mathrm{Nu}}{3 \mathrm{Pr}}\frac{c_p f_2}{\tau_d}\left({T}-T_d\right)$ is the heat transfer from the gas phase to the droplet and ${\Pi_d = \mathrm{ln}\left(1+B_M\right) \mathrm{H}\left(Y_{d,s}-Y_f\right)}$ is introduced as the mass transfer potential. $\mathrm{H}$ is a Heaviside function that prevents negative evaporation rates, as the physical model does not consider condensation.
The heat capacity of the liquid droplet is denoted by $c_l$, while $c_p$ is the heat capacity of the surrounding gas at constant pressure. $L_v$ is the latent heat of vaporization of the droplet. The Prandtl number, Pr, and Schmidt number, Sc, are evaluated using local gas transport properties. $f_2$ is a heat transfer correction factor that accounts for evaporation effects. For more information on the calculation of the Nusselt number, Nu, Sherwood number, Sh, and the two correction coefficients $f_1$ and $f_2$, the reader is referred to the work of Miller and Bellan \cite{Miller1999}. The mass transfer number is defined as $B_M = ({Y_{d,s}-{Y}_F})/({1-Y_{d,s}})$, where $Y_{d,s}$ is calculated by assuming vapor-liquid equilibrium at the liquid-gas interface. 
The "$\mathrm{\infty-rule}$" is used to approximate the effective thermodynamic and transport properties within the boundary layer.
\ifdefined\isthesis
\color{black}
\else
Besides Eq.~(\ref{eq:evapMres}), the droplet mass can change via breakup events. The Kelvin-Helmholtz Rayleigh-Taylor model is applied for secondary breakup \cite{patterson1998modeling}, while a Rosin-Rammler distribution is used for the initial droplet size to mimic the primary breakup of the liquid jet. The statistical parcel method by Dukowicz \cite{Dukowicz1980}, which pools droplets of similar properties into parcels, is used to reduce the computational effort. 
\par
Note that both breakup models are not predictive and rely on parameter calibration to provide reasonable results.
Also, note that the described drag and evaporation models have been developed for single droplets assuming fully developed boundary layers and homogeneous surrounding flow fields. Those assumptions are violated under typical CI engine injection conditions. Still, these models provided reasonable results in many studies. Consequently, either the violation of model assumptions does not drastically affect the model prediction, or modeling errors are compensated by the breakup parameter calibration. 
\fi

\subsection{Spray Subfilter Model}\label{sec:spray_sgs}
The presented spray model has been developed for Direct Numerical Simulations (DNS). The effect of velocity and scalar subfilter fluctuation on the droplet evolution is neglected in most LES and RANS studies, i.e., the filtered flow field solutions $\fav{u_j}$, $\fav{T}$, and $\fav{Y}_F$ are used in Eqs.~(\ref{eq:uDres})-(\ref{eq:evapTres}) for $u_j$, $T$, and $Y_f$, respectively. De and Kim \cite{De2013} used a stochastic model that invokes the unfiltered flow field variables by randomly sampling presumed distribution functions for $u_j$ and $Z$. 
Apart from considering flow field fluctuations on the droplet evolution, the method presented by De and Kim additionally allows closing the filtered evaporation source term in Eq.~\ref{eq:Z2trans} consistently to the local mixture FDF.
Note that this is an important model property as the droplet evolution, and hence $\ol{\dot{S}}_{\rho}$, might, arguably, be tunable via the breakup parameters, while the ratio of $\ol{\dot{S}}_{\rho}$ and $\ol{\dot{S}_{\rho} \left(2Z-Z^2\right)}$ is governed solely by the local FDF and the functional dependency of $\dot{S}_{\rho}$ on $Z$. In other words, the amount of mixture fraction variance produced by droplet evaporation for a given evaporated fuel mass is only marginally affected by the spray parameter tuning process. De and Kim applied their model to the Sidney dilute spray burner. They reported that droplet subfilter closure models yield significant differences in evaporate rates and mixing fields at locations close to the nozzle for high jet velocities and spray densities. \par
While the method presented by De and Kim models the droplet evolution and filtered evaporation source terms consistently to the subfilter distribution, the computational overhead of obtaining the subfilter distribution along all particle trajectories is considerable and increases the computational load imbalance originating from dense spray regions. Fuel injection in CI engines is typically characterized by dense spray jets, which are present only in a small fraction of the combustion chamber volume. Thus, a simpler and computationally more cost-efficient approach is employed in this study, which does not capture the effect of all subfilter fluctuations on the droplet evaporation rate but aims to model the filtered evaporation source terms consistently to the prescribed FDF. 
\par
The model used in this work assumes that the timescales of the subfilter fluctuations are smaller than the characteristic droplet timescale, i.e., the droplet state changes only slightly while experiencing the entire gas field subfilter distribution. Additionally, fluctuations of velocity and thermodynamic properties are neglected. Invoking those assumptions, Eqs.~(\ref{eq:uDres})-(\ref{eq:evapTres}) can be simplified to
\begin{align}
\tdt{u_{d,j}}&=\frac{f_1}{\tau_d}\left(\fav{u}_j - u_{d,j}\right),\label{eq:uDfil}  \\ 
\tdt{m_d} &=-\frac{\mathrm{Sh}}{3\mathrm{Sc}}\frac{m_d}{\tau_d}\ol{\Pi}_d,  \label{eq:evapMfil} \\
\tdt{T_d}& = \frac{\ol{\dot{Q}}_d}{c_l} + \frac{1}{m_d}\tdt{m_d}\frac{L_v}{c_l}.
\label{eq:evapTfil}
\end{align}
$\ol{\dot{Q}}_d = \frac{\mathrm{Nu}}{3 \mathrm{Pr}}\frac{c_p f_2}{\tau_d}\left(\ol{T}-T_d\right)$ is the filtered heat flux. The filtered temperature $\ol{T}$ is related to its Favre-filtered counterpart by $\ol{T} = \ol{\rho}\fav{\left(\frac{T}{\rho}\right)}$ or $\ol{T} = \frac{1}{\fav{(T/W)}}\fav{\left(T\frac{T}{W}\right)}$ when invoking the ideal gas law and neglecting subfilter pressure fluctuations. 
In order to model $\ol{T}$, a new temperature variable $\Delta T_\rho$ is introduced that accounts for the effect of density-weighted filtering, i.e., $\ol{T} = \fav{T} + \Delta T_{\rho}$. It is further assumed that $\Delta T_{\rho}$ is primarily affected by mixture fluctuations and hence can be well approximated using the temperature and molecular weight from the flamelet solution according to
\begin{equation}
\Delta T_{\rho} = \frac{1}{\fav{(\hat{T}/\hat{W})}}\int_{-\infty}^\infty  \ff  \hat{T} \frac{\hat{T} }{{\hat{W}}} \mathrm{d}\psi - \int_{-\infty}^\infty  \ff \hat{T} \mathrm{d}\psi,
\end{equation}
where $\fav{(\hat{T}/\hat{W})} = \int_{-\infty}^\infty  \ff \hat{T}/\hat{W} \mathrm{d}\psi $.
\par
Similar to the filtered temperature, the filtered mass transfer potential $\ol{\Pi}_d$ is also approximated using the local FDF and the flamelet solution and reads
\begin{equation}
\ol{\Pi}_d =   \frac{1}{\fav{(\hat{T}/\hat{W})}}\int_{-\infty}^\infty   \ff \Pi_d \frac{\hat{T}}{\hat{W}} \mathrm{d}\psi.\label{eq:FilTransferTerm}
\end{equation}
Conservation of mass, momentum, and energy yields the volumetric source terms in Eqs.~(\ref{eq:ContinuityFiltered})-(\ref{eq:EnergyFiltered}) as
\begin{align}
\ol{\dot{S}}_\rho = &- \sum_d^{n_p}\left[ \frac{n_{d,p}w^c_{p \rightarrow c} }{\Delta V}  \left(\tdt{m_d} \label{eq:filRhoDot}\right)\right], \\
\ol{\dot{S}}_{u_j} = & -\sum_p^{n_p}\left[ \frac{n_{d,p}w^{f_j}_{p \rightarrow c}}{\Delta V} \left(m_{d,p} \tdt{u_{d,j}} +  \tdt{m_d}u_{d,j}\right)\right], \\
\ol{\dot{S}}_{e} = & -\sum_p^{n_p}\left[ \frac{n_{d,p}w^c_{p \rightarrow c}}{\Delta V}  \left(\ol{\dot{Q}}_{d,p} + \tdt{m_d} e_{d,p}\right)\right].
\end{align}
Here, $\Delta V$ is the grid cell volume, $n_p$ is the number of parcels in the surrounding cells, $n_{d,p}$ is the number of droplets in parcel $p$, while $w^c_{p \rightarrow c}$ and $w^{f_j}_{p \rightarrow c}$ represent distribution kernels that distribute the source terms from the parcel location onto the Eulerian grid cell centers and cell faces, respectively. $e_{d,p}$ is the energy of the fuel vapor evaluated at $T_d$. 
\par
The filtered evaporation source term in the first subfilter mixture fraction moment equation (Eq.~(\ref{eq:Z1trans})) is also given by Eq.~(\ref{eq:filRhoDot}). The filtered $\fav{Z^2}$ evaporation source term can be expressed consistently to the FDF via
\begin{align}
\begin{split}
&\ol{\dot{S}_{\rho} \left(2Z-Z^2\right)} =  \\
&\quad \sum_d^{n_p} \left\{ \frac{n_{d,p} w^c_{p \rightarrow c}}{\Delta V}  \left[\frac{\mathrm{Sh}}{3\mathrm{Sc}}\frac{m_d}{\tau_d} \left(2\ol{\Pi_d Z} -\ol{\Pi_d Z^2}\right)\right]\right\},
\end{split}
\end{align}
where the filtered correlation terms are obtained from the FDF and the flamelet solution
\begin{align}
\ol{\Pi_d Z} & = \frac{1}{\fav{(\hat{T}/\hat{W})}} \int_{-\infty}^\infty   \ff\Pi_d \frac{\hat{T}}{\hat{W}} \psi  \mathrm{d}\psi \label{eq:FilTransferTerm2}, \\
\ol{\Pi_d Z^2} &=\frac{1}{\fav{(\hat{T}/\hat{W})}} \int_{-\infty}^\infty   \ff\Pi_d  \frac{\hat{T}}{\hat{W}} \psi^2 \mathrm{d}\psi\label{eq:FilTransferTerm3}.
\end{align}
Note that $Y_{d,s}$ is the only droplet-dependent parameter in Eq.~(\ref{eq:FilTransferTerm}) and Eqs.~(\ref{eq:FilTransferTerm2})-(\ref{eq:FilTransferTerm3}). In order to reduce the computational load imbalance, $\ol{\Pi}_d$ and $2\ol{\Pi_d Z}-\ol{\Pi_d Z^2}$ are tabulated over $Y_{d,s}$ in each cell before droplet integration. During droplet integration, $\ol{\Pi_d}$ and $2\ol{\Pi_d Z}-\ol{\Pi_d Z^2}$ are obtained by interpolating in space to each droplet location and interpolating in $Y_{d,s}$ direction. Note that even though the computational cost and load imbalance are reduced, the presented approach is more memory demanding.
\ifdefined\isthesis
\color{black}
\fi

\ifx\isthesis\undefined
\section{Model Application}\label{sec:application}
\else\color{blue}
\fi
The presented model is applied in simulations of the ECN Spray A case in this section. First, the case and the numerical methods are briefly described. Afterwards, the model realizability and the flow parameter predictions are discussed.
\subsection{Application Case and Simulation Setup}
The Spray A case is a high-pressure, single-hole \textit{n}-dodecane injection case defined by the Engine Combustion Network (ECN) \cite{ECN}. \begin{table}[t]
\centering
\caption{Case setup}
\begin{tabular}{lll}
\hline
 Condition & Value & Unit  \\
\hline
Nozzle					& A210677 &[-] \\
Ambient gas temperature		& 892.3 &[K]\\
Ambient gas pressure 		& 60.5 &[bar]\\
Injection pressure			& 1504 & [bar] \\
Fuel temperature			& 373 & [K] \\
\label{table:setup}
\end{tabular}
\end{table}
This study focuses on the inert case under reference conditions. Note that the flamelet solution evolution is independent of the conditional flow parameters for an inert case. Thus, there is no feedback from the conditional flow parameters to the flow, and hence, the flow parameter model predictions can be compared directly.
For more details of the case and experimental methods, the reader is referred to Pickett et al. \cite{Pickett2011}. 
The mass flow rates and initial droplet velocities are prescribed according to the nozzle specifications and experimental conditions using the injection rate model provided by CMT \cite{CMT}. 
The initial and boundary conditions are adopted from the experimental measurements (rather than the nominal values) provided on the ECN webpage \cite{ECN}. The simulation case is summarized in Tab.~\ref{table:setup}.
\par
Three numerical grids are studied, ranging from 80~$\mathrm{\mu}$m to 120~$\mathrm{\mu}$m minimum grid spacing at the nozzle orifice. The grid spacing is stretched in streamwise and cross-streamwise directions. In the streamwise direction, a growth rate of 0.2\% per cell is applied within the first 30~mm from the nozzle orifice, while a growth rate of 1.5\% is used further downstream. In the cross-streamwise directions, a growth rate of 0.2\% per cell is used within 7.5~mm distance from the spray center axis, while a growth rate of 2.5\% is applied further outside. Tab. \ref{table:grids} summarizes the numerical grids parameters.
\par
The nozzle geometry is approximated using a stair-step method, and a no-slip wall boundary condition is applied on the nozzle surface. The flamelet fuel vapor temperature boundary condition is computed from the liquid fuel temperature corrected by the latent heat of vaporization \cite{Knudsen2015}. 
The spray breakup model parameters have been calibrated to match the measured liquid and vapor penetration lengths. The initial droplet sizes follow a Rosin-Rammler distribution with the characteristic droplet diameter, $d_\mathrm{RR}=12~\mathrm{\mu m}$, and shape parameter, $k_{RR}=2$. The KHRT parameters are $B_0=0.61$, $B_1=20.0$, $C_\tau=1.0$, and $C_3=20$. 
\begin{table}[t]
\centering
\caption{Numerical grids}
\begin{tabular}{lllll}
\hline
		& Grid C 			& Grid M 			& Grid F  & Unit \\
\hline
min($\Delta x$)		& 120 			& 100 			& 80 				&[$\mathrm{\mu}$m] \\
\#grid cells  		& 19$\cdot10^6$	& 29$\cdot10^6$ 	& 50$\cdot10^6$  	&[-]\\
\#parcels	 		& 650$\cdot10^3$ 	& 500$\cdot10^3$     & 400$\cdot10^3$	&[-]\\
min($\Delta t$)		& 110 			& 90				& 70 				&[ns] \\
\end{tabular}
\label{table:grids}
\end{table}
\subsection{Numerical Methods}
All simulations were performed using the in-house flow solver CIAO. The filtered Navier-Stokes equations are solved on a structured staggered Cartesian grid. Apart from scalar convection terms, which are approximated with the WENO5 scheme \cite{liu1994}, 2\textsuperscript{nd} order central differences are employed for spatial discretization. The time integration is performed using an explicit low-storage five-stage Runge-Kutta (RK) method with 2\textsuperscript{nd} order accuracy \cite{hu1996low,stanescu19982n}. The flow time step is limited by the acoustic Courant–Friedrichs–Lewy (CFL) condition enforcing CFL numbers below 1. The Lagrangian particles are advanced using a 2\textsuperscript{nd} order RK method that employs adaptive time-stepping. 
\par
The incomplete beta function ($B_{\psi^*}$) is evaluated using a continued fraction method \cite{Cuyt2008} in a vectorized implementation that allows for better computational efficiency. The partial derivates of $\iif$, $\fav{F}_Z$, and $\fav{F}_Z^*$ with respect to $\fav{Z}$ and $\fav{Z^2}$ are obtained numerically using 2\textsuperscript{nd} order central differences, where the step size is proportional to the subfilter filter mixture fraction variance.
\subsection{Spray LES Model Validation}
\begin{figure}[t]
  \centering
  \footnotesize
  \input{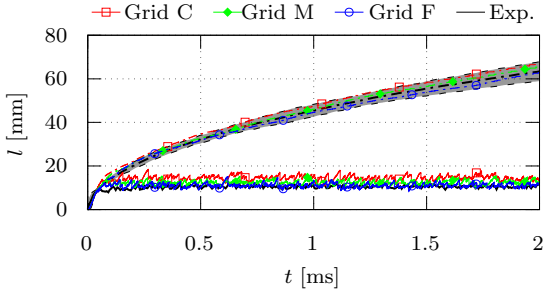}
  \caption{Spray penetration length, $l$, over time, $t$, predicted by simulations and measured in experiments \cite{Pickett2011,Pickett2015}. Liquid and vapor penetration are displayed by solid and dashed lines, respectively.}  \label{fig:penetration}
\end{figure}
Apart from spray penetration data, quantitative mixing field measurements have been performed by Pickett et al. using Rayleigh-scatter imaging \cite{Pickett2011}.  
\begin{figure}[t]
  \centering
  \footnotesize
  \input{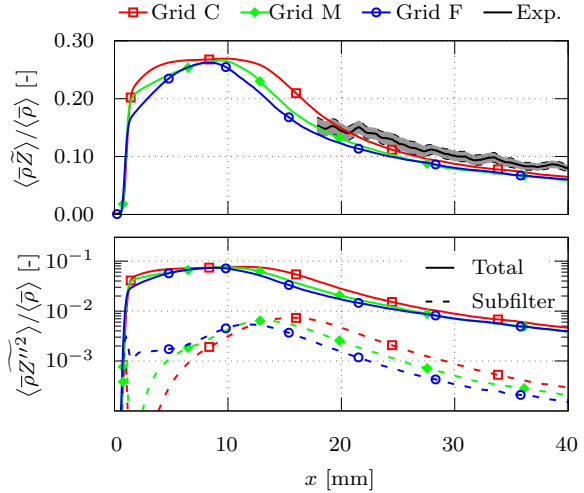}
  \caption{Mixing field statistics on the spray centerline. Top: Temporally Favre-averaged filtered mixture fraction for all numerical grids and ensemble-averaged experimental data \cite{Pickett2011}. Bottom: Temporally Favre-averaged mixture fraction variance (solid lines) and subfilter contributions (dashed lines).}  \label{fig:mixture}
\end{figure}
The LES predictions are compared to experimental results in this section.
Fig. \ref{fig:penetration} shows the liquid and vapor penetration over time obtained by the LES model for each grid in comparison to experimental results obtained from Schlieren \cite{Pickett2011} and by Diffused Back-Illumination (DBI) measurements \cite{Pickett2015}. 
The liquid penetration length (LPL) represents the farthest axial distance to the position, where the liquid volume fraction exceeds {0.1~\%} evaluated in cylindrical volumes of {1~mm} length and {1~mm} diameter. The vapor penetration is obtained from Schlieren imaging in the experiments \cite{Pickett2011}, while the numerical value represents the farthest distance to the injector orifice, where a vapor mass fraction of {0.1~\%} is exceeded. The LES model converges toward the experimental results with increasing grid resolution. The fine grid (Grid F) shows good agreement in terms of both vapor and liquid penetration. Note that only a single LES realization has been performed while the experimental vapor penetration length is ensemble-averaged. However, the shot-to-shot variations are small, as depicted by the grey area in Fig. \ref{fig:penetration} that reflects the 95\% confidence interval.
\par
The top plot in Fig.~\ref{fig:mixture} shows the averaged filtered mixture fraction on the spray centerline, where $x$ refers to the axial distance from the injector orifice. Note that ensemble-averaging has been applied to the experiments during a quasi-steady state, while single LES realizations have been performed and temporally Favre-averaged from 1~ms to 4~ms. The mixture fraction variance is shown in the bottom plot. Note that the solid lines include the variances originating from temporal averaging and the subfilter contributions. The latter are also separately plotted using dashed lines.
\begin{figure*}[t]
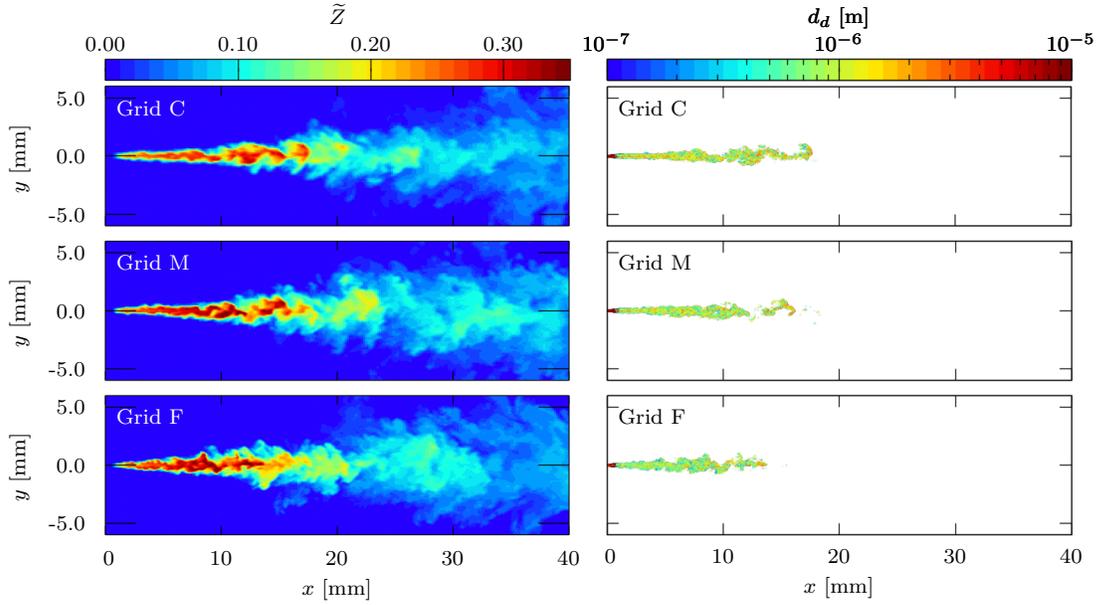

  \centering
  \footnotesize
  \ifx\isthesis\undefined
    \input{figures/mixing_fields.tex}
  \else
    \resizebox{1.025\linewidth}{!}{\input{figures/mixing_fields.tex}}
  \fi
  \caption{Instantaneous mixing fields (left) and liquid fuel parcel distribution (right) at $t=1$~ms. Grid C (top), Grid M (center), and Grid F (bottom).} \label{fig:mixing_fields}  
\end{figure*}
For all grid resolutions, three successive phases can be distinguished from the mixture fraction statistics on the centerline:
\par
The first phase is characterized by a steep gradient in both mean mixture fraction and mixture fraction variance near the nozzle orifice. Here, the fuel jet disintegrates into tiny droplets leading to small Sauter mean droplet diameters (SMD) in an unsaturated environment, and thus, rapid evaporation.
\par
The second phase shows a moderate mean mixture fraction gradient until the maximum is reached slightly before the LPL. This indicates that fuel evaporation is limited by saturation, and the entrainment of hot ambient unsaturated gas is required for further vaporization. The mixture fraction subfilter variance profiles indicate that turbulence is produced in the gas phase in this stage. Note that the total mixture fraction variance is affected by both turbulence and intermittent droplet evaporation, and hence, the whole mixture fraction variance is not a good measure of turbulence intensity. The laminar-turbulent transition length decreases with increasing grid resolution promoting air entrainment for the refined grid, thus, reducing the penetration length of liquid fuel droplets.
\par
In the third phase, most fuel droplets are evaporated and the mixture fraction field is governed by turbulent mixing as both the mean mixture fraction and the mixture fraction variances decay with increasing distance. Initially, the mean mixture fraction values are highest for the coarse grids, which is caused by a prolonged LPL. Thus, the fact that Grid C and Grid M initially show better agreement with the experimental data can be attributed to error compensation. Further downstream, all simulations approach similar values for the mixture fraction mean and variance. The Grid F case underpredicts the experimentally obtained mean mixture fraction at all positions but shows a similar slope.
\par
A qualitative comparison of the instantaneous mixture fields and liquid fuel parcel distribution at ${t=1}$~ms is given in Fig.~\ref{fig:mixing_fields}. It can be seen that the initial jet is much more coherent in the Grid C case for distances smaller than $x\approx10$~mm, while instabilities form further upstream for the Grid M and Grid F cases. As indicated by the mixing field statistics on the spray centerline, it appears that the under-resolution of the near-nozzle flow in the Grid F case causes a delayed formation of jet instabilities, resulting in overprediction of the LPL.
\subsection{LCFP Model Realizability}
\begin{figure}[t]
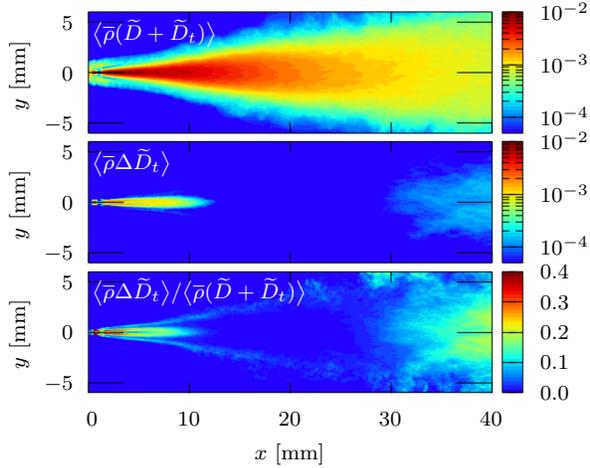

  \centering
  \footnotesize
  \ifx\isthesis\undefined
  \input{figures/realizability.tex}
  \else
  \resizebox{0.6\linewidth}{!}{\input{figures/realizability.tex}}
  \fi
  \caption{Time-averaged total diffusivity (top), eddy-diffusivity correction (center), and correction factor (bottom) for Grid F.} \label{fig:realizability}
\end{figure}
This section discusses the model realizability, i.e., whether strictly positive model outputs for the conditional dissipation rate and conditional spray source term are ensured or can be enforced.
The consistent conditional dissipation rate is given by Eq.~(\ref{eq:condCHIexact}), which consists of a sum of multiple terms. It should be noted that $\fav{\chi}_Z {\partial\iif}/{\partial{\fav{Z^2}}}$ is the only strictly positive term. All other terms can potentially become negative. Those terms scale with $\left(\fav{D}+\fav{D}_t\right)$ and external parameters that depend on the local flow field, and thus, it can be concluded that the LCFP model does not inherently enforce $\cond{\chi_Z}\geq0$. Note that $\cond{\chi_Z}<0$ occurs when either the presumed FDF shape is not able to capture the relevant physics or when the models for $\fav{D}_t$ and $\fav{\chi}_Z$ fail to provide consistent values. Hence, the realizability of the LCFP model depends strongly on the considered simulation case and employed subfilter models. Realizability is given when the ratio $\fav{\chi}_Z/\left(\fav{D}+\fav{D}_t\right)$ is sufficiently large.
Hence, a simple approach to enforce realizability is to reduce $\fav{D}_t$ by a correction term $\Delta\fav{D}_t$ in case Eq.~(\ref{eq:condCHIexact}) yields negative values for $\ff\cond{\chi_Z}$.
\par
The time-averaged total diffusivity and eddy-diffusivity correction term are shown in the top and the center plot of Fig.~\ref{fig:realizability}, respectively. The bottom plot of Fig.~\ref{fig:realizability} shows the ratio of eddy-diffusivity correction to the total uncorrected diffusivity, indicating that the correction is required mainly in three regions: Inside the dense spray region, at the vapor jet interface, and in regions far downstream, where the subfilter variances are small. Note that the dynamic model used to evaluate $\fav{D}_t$ employs the Germano identity \cite{Germano1991} that assumes the same model constant holds for two different filter widths. Due to scale similarity, this assumption is well-justified for turbulence in the inertial subrange. But it does not necessarily hold at the vapor jet interface or in the presence of dispersed droplets. Thus, it is unsurprising that the model constants need to be corrected in those regions to avoid violating physical constraints. Nevertheless, only moderate corrections are required in most regions, which are within the model uncertainties.
\par
\begin{figure*}[t]
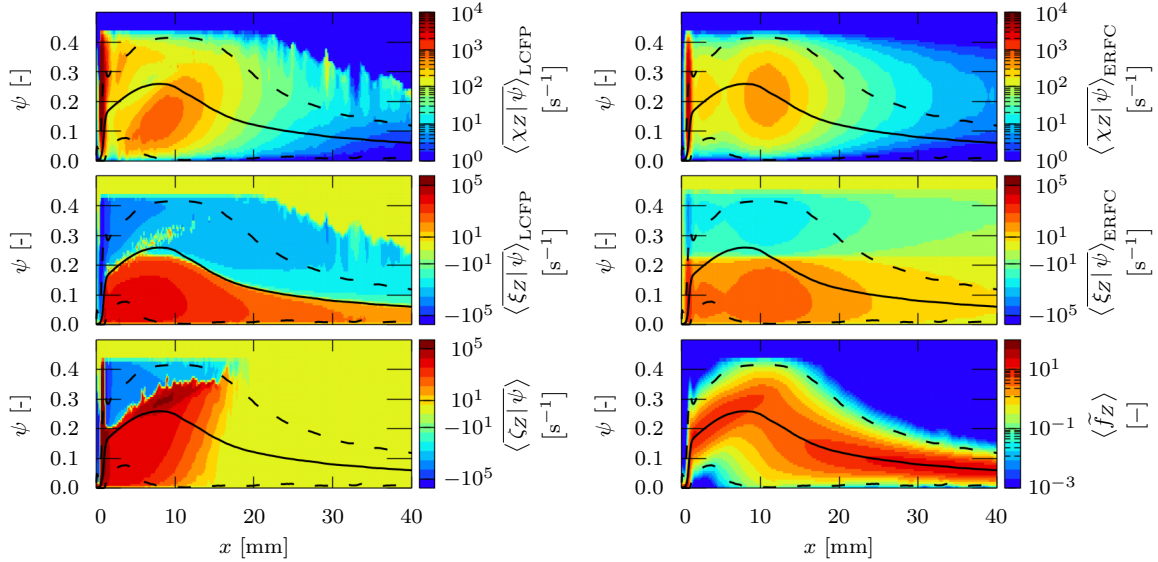

  \centering
  \footnotesize
  \ifx\isthesis\undefined
  \input{figures/condParams.tex}
  \else
  \resizebox{1.025\linewidth}{!}{\input{figures/condParams.tex}}
  \fi
  \caption{Temporally averaged conditional flow parameters on the spray centerline obtained from the LCFP model (Grid F) and the ERFC model \cite{Peters2000,Peters1984}. The local mean mixture fraction is depicted by the solid black line. The dashed black lines denote the interval with $99.8\%$ probability.}
  \label{fig:condParams}
\end{figure*}
Apart from the conditional scalar dissipation, physical constraints exist for the evaporation source term, i.e., $\cond{\zeta_Z}\geq0$ since condensation is not included in the mass transfer model. The conditional evaporation source term can be modeled consistently to the FDF evolution by Eq.~(\ref{eq:condSrc}). This equation consists of three terms, each of which is a product of filtered moment source terms and sample space-dependent functions. The sample space-dependent functions are determined by the filtered moments. The filtered moment source terms can, similarly to the flow parameters in Eq.~(\ref{eq:condCHIexact}), be viewed as external parameters. However, $\ol{\dot{S}_{\rho}}$ and $\ol{\dot{S}_{\rho} \left(2Z-Z^2\right)}$ are, in contrast to the flow parameters in Eq.~(\ref{eq:condCHIexact}), not independent of each other. More precisely, the ratio $\ol{\dot{S}_{\rho} \left(2Z-Z^2\right)}/\ol{\dot{S}_{\rho}}$ is governed by the FDF as described in section~\ref{sec:spray_sgs}. Thus, the realizability of Eq.~(\ref{eq:condSrc}) is solely governed by the presumed FDF shape and mixture fraction moments, and thus, the model cannot be tweaked to enforce the physical constraint. In other words, when Eq.~(\ref{eq:condSrc}) predicts a negative conditional spray source term, it can be concluded that the selected FDF model is inappropriate to capture the relevant physics. 
\subsection{Local Conditional Flow Parameters}
The model predictions for the conditional flow parameters are presented in this section. 
Fig.~\ref{fig:condParams} shows the time-averaged conditional flow parameters of the fine grid case (Grid F) along with the time-averaged FDF on the spray centerline. The flow parameters obtained by the widely applied inverse-error function model \cite{Peters2000,Peters1984}, which will be referred to as the ERFC model in the following, are also plotted as a reference. 
\par
The conventional model neglects curvature effects allowing to obtain $\cond{\xi_{Z}}_\mathrm{ERFC}$ from $\cond{\chi_{Z}}_\mathrm{ERFC}$ via
${\cond{\xi_{Z}}_\mathrm{ERFC}= \frac{1}{4\hat{\rho}}\left[\dPsi{\hat{\rho}\cond{\chi_Z}_\mathrm{ERFC}} + \frac{\cond{\chi_Z}_\mathrm{ERFC}}{\hat{D}} \dPsi{\hat{\rho}\hat{D}}\right]}$.
The solid black lines in Fig.~\ref{fig:condParams} correspond to the mean mixture fraction, $\avg{\fav{Z}}$, while the dashed lines denote the mixture interval around the mean with $99.8\%$ probability. 
\par
The spatial evolution of the temporally averaged FDF, $\avg{\ff}$, is shown in the bottom right plot of Fig.~\ref{fig:condParams}. It gives the first insight into the mixture formation process. The mixture distribution starts with an almost pure oxidizer at the nozzle orifice for $x=0$~mm due to the large SMD of the initial droplets and the corresponding low evaporation rate. At $x \approx 1$~mm, a rapid change in the FDF is observed. The initial droplets disintegrate at this position due to Rayleigh-Taylor breakup, which drastically increases the SMD, thus, resulting in fast evaporation until the mixture is saturated. After reaching the saturated state, the evaporation rate is controlled by droplet heating, i.e., raising droplet temperatures increase the fuel vapor mass fraction on the droplet surface, allowing the droplets to evaporate further ($1$~mm~$\lesssim x \lesssim 6$~mm).  At $x\approx 6$~mm, it can be seen that the first pure oxidizer is detected again in $\avg{\ff}$, i.e., the jet instabilities have grown sufficiently to break up the coherent core and entrain pure oxidizer into the spray centerline. The remaining droplets evaporate rapidly from this point on. Beyond the LPL, the mixture distribution evolves similarly to a turbulent jet, i.e., the mixture fluctuations decay, and the distribution becomes increasingly narrow with increasing distance. 
\par
The temporally averaged conditional dissipation rates predicted by the LCFP and the ERFC models are shown in the top left and the top right plot of Fig.~\ref{fig:condParams}, respectively. Significant differences are observed for $2\lesssim x \lesssim15$~mm. 
The shape of the LCFP model predicts an unsymmetric, bimodal dissipation rate profile and shows larger peak dissipation rates for mixtures fractions below $\avg{\fav{Z}}$. Hence, the LCFP model promotes mixing in leaner mixtures compared to the ERFC model.
Beyond the LPL, both models yield remarkably similar results in regions where the FDF is sufficiently large.
Consequently, conservation errors can be expected to be small if the conventional ERFC model is applied together with a modified beta distribution provided that the same mixture fraction bounds are used for $\ff$ and $\cond{\chi_Z}_\mathrm{ERFC}$. Hence, it is unsurprising that flamelet models, which employ the ERFC model, produce good predictions of reactive unity Lewis number scalars in many studies. 
\par
The time-averaged conditional diffusion rate predictions of both models are compared in the second row of Fig.~\ref{fig:condParams}. Note that to properly depict $\avg{\cond{\xi_Z}}$, which varies within several orders of magnitude and can take both positive and negative values, the transfer function, ${t_c=\mathrm{sign}(\avg{\cond{\xi_Z}})\mathrm{log}(1+|\avg{\cond{\xi_Z}}|)}$, has been employed for the color scheme. Both models predict wave-like profiles in mixture fraction space, where negative diffusion rates appear on the rich side, while the diffusion rates are positive on the lean side. However, the magnitude and root are significantly different at almost all locations. The ERFC model enforces a symmetric profile and predicts a smaller magnitude due to the negligence of curvature effects. In contrast, the root predicted by the LCFP model closely follows the evolution of $\avg{\fav{Z}}$. 
Note that this feature is essential to preserve the mass of non-unity Lewis number scalars. Considering a non-diffusive scalar, i.e., infinity Lewis number, such as large soot particles, it can be seen from Eq.~(\ref{eq:turb_spray_flamelet}) that, in the absence of droplet sources, $\cond{\xi_Z}$ is the only contributor to the convection velocity of $\hat{\phi}$ in $\psi$-space. Thus, the root of the wave-like $\cond{\xi_Z}$ profile can be regarded a stagnation plane for soot in $\psi$-space. Consequently, all scalar mass will eventually be trapped at this stagnation plane. Since the mixture distribution becomes leaner and increasingly narrow with increasing distance, the entire scalar mass will eventually be removed from physical space if the flow parameter model predicts a ${\cond{\xi_Z}}$ with constant root. However, using ${\cond{\xi_Z}}$ of the LCFP model enforces that the scalar is transported consistently to the FDF evolution. This becomes clear when considering the limit of $x\rightarrow\infty$, i.e.,  all mixture fluctuations are decayed. Until this point, the entire scalar mass has to be transported towards $\avg{\fav{Z}}$ in order to preserve scalar mass, which is ensured by the LCFP model.
\par
While the ERFC model approximates the turbulent transport of unity Lewis number scalars in mixture fraction space without introducing significant conservation errors, it can be concluded that it inherently fails to preserve the conservation of mass of non-unity Lewis number scalars. In fact, non-premixed flamelet models struggled in the past to predict preferential diffusion effects in turbulent flames. Assuming unity Lewis numbers often resulted in better model performance, even when compared to Direct Numerical Simulations (DNS) that employed mixture-averaged diffusion models \cite{Attili2016}. The comparisons have been made using counterflow diffusion flames. The counterflow diffusion flame configuration is almost perfectly approximated by the ERFC model parameters, which are, as shown above, well-suited to predict turbulent transport of unity Lewis number species but are prone to scalar mass conservation errors when applied to preferential diffusion processes in turbulent environments.
\par
The conditional normalized evaporation source term is shown in the bottom left plot of Fig.~\ref{fig:condParams}. Again, $t_c$ has been applied to depict the entire range $\avg{\cond{\zeta_Z}}$. Most importantly, it shows that Eq.~(\ref{eq:condSrc}) does not preserve the physical limits of the evaporation model when the modified beta function is employed as negative values are obtained for $\avg{\cond{\zeta_Z}}$ in rich mixtures at many locations. This might not be surprising since, in the presence of evaporation, the mixture fraction transport equation resembles the same mathematical form as the progress variable transport equation, for which it is well known that a beta distribution fails to capture the subfilter distribution evolution. At the same time, it should be noted that scalar mass conservation will be violated if a different model is employed for $\avg{\cond{\zeta_Z}}$ together with a modified beta distribution FDF. Finding a distribution function that inherently ensures realizable conditional evaporation source terms is, however, beyond the scope of this paper and a topic for future research.
\section{Summary and Conclusion}\label{sec:conclusion}
The local conditional flow parameters, which are required for mixture fraction based turbulent combustion models, have been formulated consistently to the evolution of a presumed subfilter density function (FDF) evolution. The realizability of the model equations has been discussed. Realizable conditional dissipation rates can be enforced by manipulating the turbulent eddy-diffusivity, while the consistent conditional evaporation source term equation contains no free model parameter that could be adjusted to enforce realizability. The new model has been applied to an LES of the inert ECN "Spray A" case. The effect of grid resolution on the mixing field has been discussed, and the spray model predictions have been compared with experimental data. Good agreement has been achieved for liquid and vapor spray penetration, while the simulation underpredicts the mean mixture fraction on the spray centerline. The necessity of model corrections for achieving realizability has been investigated afterwards. A moderate correction of the eddy-diffusivity is required in some regions to obtain realizable conditional dissipation rates. Lastly, the new consistent flow parameter model (LCFP) predictions have been compared to those of the conventional inverse-error function (ERFC) model. The main conclusions are:
\begin{itemize}
\item The LCFP model predicts a bimodal conditional dissipation rate distribution in regions where liquid droplets are present and promotes mixing in lean mixtures in those regions compared to the conventional ERFC model. The LCFP and ERFC models yield similar conditional dissipation rate, $\avg{\cond{\chi_Z}}$, values downstream of the liquid spray penetration length, i.e., scalar mass conservation errors are expected to be small for unity Lewis number scalars when the ERFC model is employed for similar configurations.
\item The conditional diffusion rate, $\avg{\cond{\xi_Z}}$, predictions differ significantly in both shape and magnitude. $\cond{\xi_Z}$ contributes to the convection speed in mixture fraction space for non-unity Lewis number scalars, and hence, in the absence of droplets, the root of the $\cond{\xi_Z}$ profile in mixture fraction space represents a stagnation plane. This stagnation plane attracts scalars with $\mathrm{Le}_\phi>1$ and repels scalars with $\mathrm{Le}_\phi<1$. The roots of the ${\cond{\xi_Z}}_\mathrm{ERFC}$ profiles remain at the same mixture fraction position for all axial locations, which can cause severe scalar mass conservation errors. The root of ${\cond{\xi_Z}}_\mathrm{LCFP}$ follows the mean mixture fraction such that scalar mass is conserved during turbulent mixing. 
\item The predicted conditional normalized evaporation source term, $\cond{\zeta_Z}$, does not preserve physical contraints of pure evaporation. The model input parameters cannot be manipulated to enforce $\cond{\zeta_Z}\geq0$ without sacrificing other model consistency constraints. Hence, the modified beta function with constant mixture fraction bounds cannot capture the relevant physics under the considered conditions.
\end{itemize}
Despite these results, the impact of the model revision on the prediction of combustion and pollutant formation under reactive conditions remains to be investigated in future work.
\ifx\isthesis\undefined\else\color{black}\fi

\section*{Declaration of Competing Interest}
The authors declare that they have no known competing financial interests or personal relationships that could have appeared to influence the work reported in this paper.
\section*{Acknowledgments}
This work was funded by the Deutsche Forschungsgemeinschaft (DFG, German Research Foundation) under Germany’s Excellence Strategy – Exzellenzcluster 2186 “The Fuel Science Center” ID: 390919832. \par
The authors gratefully acknowledge the computing time provided to them at the NHR Center NHR4CES at RWTH Aachen University (project number p0020437). This is funded by the Federal Ministry of Education and Research, and the state governments participating on the basis of the resolutions of the GWK for national high performance computing at universities (www.nhr-verein.de/unsere-partner).
\section*{Supplementary material}
A detailed mathematical derivation of the turbulent non-premixed flamelet and LCFP model equations is provided in the supplementary material.


\bibliographystyle{elsarticle-num}
\bibliography{biblio}

\begin{thebibliography}{10}
\expandafter\ifx\csname url\endcsname\relax
  \def\url#1{\texttt{#1}}\fi
\expandafter\ifx\csname urlprefix\endcsname\relax\def\urlprefix{URL }\fi
\expandafter\ifx\csname href\endcsname\relax
  \def\href#1#2{#2} \def\path#1{#1}\fi

\bibitem{Tsai1995}
K.~Tsai, R.~O. Fox, Modeling multiple reactive scalar mixing with the
  generalized {IEM} model, Physics of Fluids 7~(11) (1995) 2820--2830.
\newblock \href {https://doi.org/10.1063/1.868660}
  {\path{doi:10.1063/1.868660}}.

\bibitem{Davidovic2022}
M.~Davidovic, H.~Pitsch, Formulation and importance of conservative transport
  in non-premixed flamelet models, Proceedings of the Combustion Institute
  39~(2) (2023) 2429--2438.
\newblock \href {https://doi.org/https://doi.org/10.1016/j.proci.2022.07.149}
  {\path{doi:https://doi.org/10.1016/j.proci.2022.07.149}}.

\bibitem{Fox2020}
R.~O. Fox, Effect of the conditional scalar dissipation rate in the conditional
  moment closure, Physics of Fluids 32~(11) (2020) 115118.
\newblock \href {https://doi.org/10.1063/5.0030092}
  {\path{doi:10.1063/5.0030092}}.

\bibitem{Ilgun2021proci}
A.~Ilgun, A.~Passalacqua, R.~Fox, Application of quadrature-based moment
  methods to the conditional moment closure, Proceedings of the Combustion
  Institute 38~(2) (2021) 2749--2757.
\newblock \href {https://doi.org/10.1016/j.proci.2020.07.075}
  {\path{doi:10.1016/j.proci.2020.07.075}}.

\bibitem{Cook1994}
A.~W. Cook, J.~J. Riley, {A subgrid model for equilibrium chemistry in
  turbulent flows}, Physics of Fluids 6~(8) (1994) 2868--2870.
\newblock \href {https://doi.org/10.1063/1.868111}
  {\path{doi:10.1063/1.868111}}.

\bibitem{Jimenez1997}
J.~Jim{\'{e}}nez, A.~Li{\~{n}}{\'{a}}n, M.~M. Rogers, F.~J. Higuera, A priori
  testing of subgrid models for chemically reacting non-premixed turbulent
  shear flows, Journal of Fluid Mechanics 349 (1997) 149--171.
\newblock \href {https://doi.org/10.1017/s0022112097006733}
  {\path{doi:10.1017/s0022112097006733}}.

\bibitem{Wall2000}
C.~Wall, B.~J. Boersma, P.~Moin, An evaluation of the assumed beta probability
  density function subgrid-scale model for large eddy simulation of
  nonpremixed, turbulent combustion with heat release, Physics of Fluids
  12~(10) (2000) 2522.
\newblock \href {https://doi.org/10.1063/1.1287911}
  {\path{doi:10.1063/1.1287911}}.

\bibitem{OBrien1991}
E.~E. O'Brien, T.-L. Jiang, The conditional dissipation rate of an initially
  binary scalar in homogeneous turbulence, Physics of Fluids A: Fluid Dynamics
  3~(12) (1991) 3121--3123.
\newblock \href {https://doi.org/10.1063/1.858127}
  {\path{doi:10.1063/1.858127}}.

\bibitem{Tsai1998}
K.~Tsai, R.~O. Fox, The {BMC}/{GIEM} model for micromixing in non-premixed
  turbulent reacting flows, Industrial {\&} Engineering Chemistry Research
  37~(6) (1998) 2131--2141.
\newblock \href {https://doi.org/10.1021/ie970589j}
  {\path{doi:10.1021/ie970589j}}.

\bibitem{Pope2000}
S.~B. Pope, Turbulent Flows, Cambridge University Press, 2000.

\bibitem{Reveillon2000}
J.~R{\'e}veillon, L.~Vervisch, Spray vaporization in nonpremixed turbulent
  combustion modeling: a single droplet model, Combustion and Flame 121~(1-2)
  (2000) 75--90.
\newblock \href {https://doi.org/10.1016/S0010-2180(99)00157-1}
  {\path{doi:10.1016/S0010-2180(99)00157-1}}.

\bibitem{Ge2006}
H.-W. Ge, E.~Gutheil, Probability density function ({PDF}) simulation of
  turbulent spray flows, Atomization and Sprays 16~(5) (2006) 531--542.
\newblock \href {https://doi.org/10.1615/atomizspr.v16.i5.40}
  {\path{doi:10.1615/atomizspr.v16.i5.40}}.

\bibitem{Luo2011}
K.~Luo, H.~Pitsch, M.~Pai, O.~Desjardins, Direct numerical simulations and
  analysis of three-dimensional n-heptane spray flames in a model swirl
  combustor, Proceedings of the Combustion Institute 33~(2) (2011) 2143--2152.
\newblock \href {https://doi.org/10.1016/j.proci.2010.06.077}
  {\path{doi:10.1016/j.proci.2010.06.077}}.

\bibitem{Cuyt2008}
A.~Cuyt, V.~B. Petersen, B.~Verdonk, H.~Waadeland, W.~B. Jones, Handbook of
  Continued Fractions for Special Functions, Springer Science and Business
  Media, 2008.

\bibitem{Germano1991}
M.~Germano, U.~Piomelli, P.~Moin, W.~H. Cabot, A dynamic subgrid-scale eddy
  viscosity model, Physics of Fluids A: Fluid Dynamics 3~(7) (1991) 1760--1765.
\newblock \href {https://doi.org/10.1063/1.857955}
  {\path{doi:10.1063/1.857955}}.

\bibitem{Meneveau1996}
C.~Meneveau, T.~S. Lund, W.~H. Cabot, A lagrangian dynamic subgrid-scale model
  of turbulence, Journal of Fluid Mechanics 319~(-1) (1996) 353.
\newblock \href {https://doi.org/10.1017/s0022112096007379}
  {\path{doi:10.1017/s0022112096007379}}.

\bibitem{Pierce1998}
C.~D. Pierce, P.~Moin, A dynamic model for subgrid-scale variance and
  dissipation rate of a conserved scalar, Physics of Fluids 10~(12) (1998)
  3041--3044.
\newblock \href {https://doi.org/10.1063/1.869832}
  {\path{doi:10.1063/1.869832}}.

\bibitem{Peters2000}
N.~Peters, Turbulent Combustion, Cambridge University Press, 2000.
\newblock \href {https://doi.org/10.1017/cbo9780511612701}
  {\path{doi:10.1017/cbo9780511612701}}.

\bibitem{Raman2006}
V.~Raman, H.~Pitsch, R.~O. Fox, Eulerian transported probability density
  function sub-filter model for large-eddy simulations of turbulent combustion,
  Combustion Theory and Modelling 10~(3) (2006) 439--458.
\newblock \href {https://doi.org/10.1080/13647830500460474}
  {\path{doi:10.1080/13647830500460474}}.

\bibitem{Ihme2008b}
M.~Ihme, H.~Pitsch, Prediction of extinction and reignition in nonpremixed
  turbulent flames using a flamelet/progress variable model: 2. application in
  les of sandia flames d and e, Combustion and Flame 155~(1) (2008) 90--107.
\newblock \href
  {https://doi.org/https://doi.org/10.1016/j.combustflame.2008.04.015}
  {\path{doi:https://doi.org/10.1016/j.combustflame.2008.04.015}}.

\bibitem{Kaul2011}
C.~M. Kaul, V.~Raman, A posteriori analysis of numerical errors in subfilter
  scalar variance modeling for large eddy simulation, Physics of Fluids 23~(3)
  (2011) 035102.
\newblock \href {https://doi.org/10.1063/1.3556097}
  {\path{doi:10.1063/1.3556097}}.

\bibitem{De2013}
S.~De, S.~H. Kim, Large eddy simulation of dilute reacting sprays: Droplet
  evaporation and scalar mixing, Combustion and Flame 160~(10) (2013)
  2048--2066.
\newblock \href {https://doi.org/10.1016/j.combustflame.2013.04.024}
  {\path{doi:10.1016/j.combustflame.2013.04.024}}.

\bibitem{Miller1999}
R.~S. Miller, J.~Bellan, {D}irect {N}umerical {S}imulation of a confined
  three-dimensional gas mixing layer with one evaporating
  hydrocarbon-droplet-laden stream, Journal of Fluid Mechanics 384 (1999)
  293--338.
\newblock \href {https://doi.org/10.1063/1.870271}
  {\path{doi:10.1063/1.870271}}.

\bibitem{patterson1998modeling}
M.~A. Patterson, R.~D. Reitz, Modeling the effects of fuel spray
  characteristics on diesel engine combustion and emission, SAE transactions
  (1998) 27--43\href {https://doi.org/10.4271/980131}
  {\path{doi:10.4271/980131}}.

\bibitem{Dukowicz1980}
J.~K. Dukowicz, A particle-fluid numerical nodel for liquid sprays, Journal of
  Computational Physics 35~(2) (1980) 229--253.
\newblock \href {https://doi.org/10.1016/0021-9991(80)90087-X}
  {\path{doi:10.1016/0021-9991(80)90087-X}}.

\bibitem{ECN}
\href{https://ecn.sandia.gov}{{The Engine Combustion Network (ECN)}} [cited
  2019].
\newline\urlprefix\url{https://ecn.sandia.gov}

\bibitem{Pickett2011}
L.~M. Pickett, J.~Manin, C.~L. Genzale, D.~L. Siebers, M.~P.~B. Musculus, C.~A.
  Idicheria, Relationship between diesel fuel spray vapor
  penetration/dispersion and local fuel mixture fraction, {SAE} International
  Journal of Engines 4~(1) (2011) 764--799.
\newblock \href {https://doi.org/10.4271/2011-01-0686}
  {\path{doi:10.4271/2011-01-0686}}.

\bibitem{CMT}
\href{https://www.cmt.upv.es/#/ecn/download/InjectionRateGenerator/InjectionRateGenerator}{{CMT
  Injection Rate Generator}}.
\newline\urlprefix\url{https://www.cmt.upv.es/#/ecn/download/InjectionRateGenerator/InjectionRateGenerator}

\bibitem{Knudsen2015}
E.~Knudsen, Shashank, H.~Pitsch, Modeling partially premixed combustion
  behavior in multiphase {LES}, Combustion and Flame 162~(1) (2015) 159--180.
\newblock \href {https://doi.org/10.1016/j.combustflame.2014.07.013}
  {\path{doi:10.1016/j.combustflame.2014.07.013}}.

\bibitem{liu1994}
X.-D. Liu, S.~Osher, T.~Chan, Weighted essentially non-oscillatory schemes,
  Journal of Computational Physics 115~(1) (1994) 200 -- 212.
\newblock \href {https://doi.org/10.1006/jcph.1994.1187}
  {\path{doi:10.1006/jcph.1994.1187}}.

\bibitem{hu1996low}
F.~Hu, M.~Hussaini, J.~Manthey, Low-dissipation and low-dispersion
  {Runge-Kutta} schemes for computational acoustics, Journal of Computational
  Physics 124~(1) (1996) 177--191.
\newblock \href {https://doi.org/10.1006/jcph.1996.0052}
  {\path{doi:10.1006/jcph.1996.0052}}.

\bibitem{stanescu19982n}
D.~Stanescu, W.~Habashi, 2n-storage low dissipation and dispersion
  {Runge-Kutta} schemes for computational acoustics, Journal of Computational
  Physics 143~(2) (1998) 674--681.
\newblock \href {https://doi.org/10.1006/jcph.1998.5986}
  {\path{doi:10.1006/jcph.1998.5986}}.

\bibitem{Pickett2015}
L.~M. Pickett, C.~L. Genzale, J.~Manin, Uncertainty quantification for liquid
  penetration of evaporating sprays at {D}iesel-like conditions, Atomization
  and Sprays 25~(5) (2015) 425--452.
\newblock \href {https://doi.org/10.1615/atomizspr.2015010618}
  {\path{doi:10.1615/atomizspr.2015010618}}.

\bibitem{Peters1984}
N.~Peters, Laminar diffusion flamelet models in non-premixed turbulent
  combustion, Progress in Energy and Combustion Science 10~(3) (1984) 319--339.
\newblock \href {https://doi.org/10.1016/0360-1285(84)90114-X}
  {\path{doi:10.1016/0360-1285(84)90114-X}}.

\bibitem{Attili2016}
A.~Attili, F.~Bisetti, M.~E. Mueller, H.~Pitsch, Effects of non-unity lewis
  number of gas-phase species in turbulent nonpremixed sooting flames,
  Combustion and Flame 166 (2016) 192--202.
\newblock \href {https://doi.org/10.1016/j.combustflame.2016.01.018}
  {\path{doi:10.1016/j.combustflame.2016.01.018}}.

\end{thebibliography}
%

\end{document}